\documentclass[11pt,a4paper]{article}

\usepackage{jheppub}
\usepackage[usenames,dvipsnames]{xcolor}
\usepackage{amssymb} 
\usepackage{MnSymbol}
\usepackage{pgfplots}
\pgfplotsset{compat=1.17}
\usetikzlibrary{pgfplots.fillbetween}
\usepackage{amsmath}
\usepackage{mathtools}
\usepackage{amsfonts}  
\usepackage{dsfont}
\usepackage{pdfpages}
\usepackage{verbatim}
\usepackage{stmaryrd}
\usepackage{graphicx}
\usepackage{tabularx}        
\usetikzlibrary{positioning}
\usepackage{hhline}
\usepackage{multirow}
\usepackage[most]{tcolorbox}

\usepackage{tensor}
\usepackage{mathrsfs}
\usepackage{textgreek} 
\usepackage{euscript}
\usepackage[smalltableaux]{ytableau}
\ytableausetup{centertableaux}
\usepackage{tikz}
\usetikzlibrary{decorations.pathreplacing, decorations.markings,calc,shapes.misc,decorations.pathmorphing,patterns.meta, math}
\usetikzlibrary{decorations.pathmorphing}
\usetikzlibrary{decorations.markings}
\usepackage{slashed}
\usepackage{datetime}
\usepackage{hyperref}
\usepackage{braket}
\hypersetup{
    pdfencoding=unicode,
	colorlinks=true,
	urlcolor=Maroon,
	linkcolor=RoyalBlue,
	citecolor=Maroon,
	pdftitle={dS$^4$ Metamorphosis},
	pdfauthor={Dionysios Anninos, Chiara Baracco, Vasileios Letsios, Beatrix M\"uhlmann},
	pdfdisplaydoctitle=true,
	pdfstartview=FitH,
	linktocpage=true
}
\usepackage{pgfplots}
\pgfplotsset{compat=1.17}

\usetikzlibrary{shapes}
\usetikzlibrary{arrows}
\usetikzlibrary{decorations.pathmorphing}
\usetikzlibrary{decorations.markings}
\usetikzlibrary{shapes.misc}
\tikzset{snake it/.style={decorate, decoration=snake}}
\tikzset{cross/.style={cross out, draw=black, minimum size=2*(#1-\pgflinewidth), inner sep=0pt, outer sep=0pt},
cross/.default={1pt}}
\usetikzlibrary{shapes.geometric}
\tikzset{
    partial ellipse/.style args={#1:#2:#3}{
        insert path={+ (#1:#3) arc (#1:#2:#3)}
    }
}

\usepackage{subcaption}

\newcommand{\ba}{\begin{align}}

\newcommand{\be}{\begin{equation}}
\newcommand{\ee}{\end{equation}}
\def\bd{\begin{tikzpicture}}
\def\ed{\end{tikzpicture}}

\DeclareMathOperator\tr{tr}

\allowdisplaybreaks

\def\XXint#1#2#3{{\setbox0=\hbox{$#1{#2#3}{\int}$}
     \vcenter{\hbox{$#2#3$}}\kern-.5\wd0}}

\definecolor{light-gray}{gray}{0.75}

\renewcommand\d{\text{d}}

\newcommand{\e}{\mathrm{e}}

\newcommand{\ii}{\mathrm{i}}

\renewcommand{\ge}{\geqslant}

\renewcommand{\geq}{\geqslant}

\definecolor{bleudefrance}{rgb}{0.19, 0.55, 0.91}

\definecolor{vert}{rgb}{0.1367 0.543 0.1367}

\definecolor{pink}{rgb}{1.0, 0.13, 0.32}

\begin{document}

\vspace*{2.5cm}
\begin{center}
{\LARGE \textbf{\textsc{dS$^4$ Metamorphosis}}}

\vspace*{1.7cm}

{\bf
\mbox{Dionysios Anninos$^{4,2}$, 
Chiara Baracco$^{4}$, 
Vasileios A. Letsios$^{3}$, 
Beatrix M\"uhlmann$^{1}$}
}

\vspace*{0.6cm}

{\footnotesize
$^1$ School of Natural Sciences, Institute for Advanced Study, Princeton, NJ 08540, USA\\
$^2$ Instituut voor Theoretische Fysica, KU Leuven, Celestijnenlaan 200D, B-3001 Leuven, BE\\
$^3$ Physique de l’Univers, Champs et Gravitation, UMONS, Place du Parc 20, 7000 Mons, BE\\
$^4$ Department of Mathematics, King’s College London, The Strand, London WC2R 2LS, UK
}

\vspace*{0.4cm}

{\footnotesize\textsf{
dionysios.anninos@kcl.ac.uk, 
chiara.baracco@kcl.ac.uk, 
vasileios.letsios@umons.ac.be, 
beatrix@ias.edu
}}

\vspace*{0.6cm}
\end{center}

\vspace*{1.5cm}
\noindent
\begin{abstract}\noindent
We study the Euclidean path integral of higher spin gravity on $S^4$. Based on a one-loop analysis, we are led to a gluing formula expressing the $S^4$ path integral in terms of an underlying $S^3$ path integral. We view the three-sphere as a boundary hypersurface splitting the four-sphere into two halves. For a higher spin spectrum containing even spins only, the resulting boundary theory living on the $S^3$ cut is the $\mathrm{Sp}(N)$ invariant sector of $N\in \mathbb{Z}^+$ anti-commuting, conformally coupled free scalars, with conformal higher spin sources mediating the gluing. This boundary $\mathrm{Sp}(N)$ theory was previously shown to compute the Hartle-Hawking wavefunction  at $\mathcal{I}^+$ in the higher spin dS$_4$/CFT$_3$ correspondence. In contrast to the infinite spatial volume of $\mathcal{I}^+$, here the conformal fields populate a finite size $S^3$ hypersurface of $S^4$. For theories with both bosonic and fermionic higher spin fields,  the gluing formula is instead built from an $\mathcal{N}=2$ superconformal boundary field theory coupled to $U(N)$ invariant superconformal sources. Under this assumption, the leading contribution to the four-sphere partition function is $2^N$, and we observe exact cancellations at one-loop. 
\end{abstract}

\newpage

\tableofcontents

\section{Introduction}

The character of quantum information coding a cosmological spacetime, including our own Universe, is an important open question. Lessons from black holes \cite{Bekenstein:1973ur,Hawking:1975vcx,Ryu:2006bv,Penington:2019npb} suggest that the underlying information, no matter its precise nature, is distributed in a surprising and strikingly non-local way, contrasting the seemingly ordinary local organisation characteristic of semiclassical phenomena. Identifying its underlying structure may play an important role in clarifying how low-energy effective field theory breaks down, and in determining the degrees of freedom that replace the semiclassical description of cosmology near a spacelike singularity. Another lesson from black holes is that a quantitative treatment of such physical information  may require a significant and often uncomfortable departure \cite{Strominger:1996sh,Saad:2019lba} from any ordinary looking physical system. In parallel, it is important to identify sharp macroscopic diagnostics that are reliably calculable within the semiclassical regime, in order to constrain any putative microphysical perspective. In the context of the quantum theory of black holes, this aspect of the problem is suitably addressed by black hole thermodynamics as computed from the Euclidean gravitational path integral \cite{Gibbons:1976ue}, which produces quantitative thermodynamic formulae for a wide variety of black holes.
\newline\newline
Less is known about Euclidean quantum gravity in the cosmological setting, and the construction of microphysically complete cosmological theories. Nonetheless, it was proposed by Gibbons and Hawking that the logarithm of the gravitational path integral   on a sphere \cite{Gibbons:1976ue,Gibbons:1977mu,Anninos:2020hfj,Muhlmann:2022duj} for a theory with $\Lambda>0$, namely  $\mathcal{S} = \log \mathcal{Z}_{\text{grav}}$, computes the quantum entropy of the de Sitter horizon. Semiclassically, $\mathcal{S}$ scales as $\tfrac{1}{G_N \Lambda}$.  Technically speaking, $\mathcal{Z}_{\text{grav}}$  has the virtue of being both a gauge invariant and locally field redefinition invariant function of the parameters of the theory. Moreover, Hartle and Hawking \cite{Hartle:1983ai} argue that the Euclidean path integral  constructs an interesting solution to the Wheeler-DeWitt equation. Mathematically complete models of four-dimensional de Sitter space under theoretical control are less abundant, though exceptions to this rule have been pursued for higher spin theories \cite{Anninos:2011ui,Anninos:2017eib}, and theories with boundaries that rearrange the AdS$_4$/CFT$_3$ Hilbert space into a de Sitter one \cite{Silverstein:2024xnr}.
\newline\newline
In what follows, we consider exotic theories of cosmological spacetimes that admit four-dimensional semiclassical de Sitter solutions. Their exotic character arises from the presence of an infinite tower of non-linearly interacting massless higher spin gauge fields \cite{Vasiliev:1990en}. The spectrum includes a massless spin-two gauge field, identified in the free approximation with the linearised graviton, which becomes interacting at higher orders. The exotic nature of these theories leads to a rich and constraining mathematical structure, as they are subject to an infinite extension of the diffeomorphism group. In the presence of this structure, such theories likely do not suffer from the usual non-renormaliseablility issues of the standard gravitational effective field theory. The Lorentzian version of these higher spin theories has been previously discussed in the context of a higher spin dS$_4$/CFT$_3$ correspondence \cite{Anninos:2011ui}, where one anchors information at the infinite future $\mathcal{I}^+$ and builds something akin to a wavefunctional of the higher spin world in terms of a dual rank-$N$ vector-like conformal theory. In Euclidean signature the higher spin fields reside on a four-dimensional fluctuating sphere, void of  external information tied to a boundary of any sort. 
\newline\newline
It is not a priori clear how the Euclidean theory on $S^4$ should be related, if at all, to the Lorentzian dS$_4$/CFT$_3$ perspective which naturally resides at $\mathcal{I}^+$.\footnote{This is to be contrasted to a quantum field theory in a non-dynamical de Sitter background, where the Euclidean picture is directly related to the computation of physical expectation values in the Euclidean vacuum.} A microscopic Hilbert space describing the quantum completion of the classical higher spin Lorentzian dS$_4$ theory was proposed in \cite{Anninos:2011ui}. It was observed, following the lines of \cite{Higuchi:1991tk}, that the physical Hilbert space should be invariant under the entire residual higher spin isometry group, which is an infinite extension of the $\mathrm{SO}(1,4)$ de Sitter isometry group. Upon implementing these constraints, and taking into account that the microscopic operators are built from $N$-component fields at each spatial point on $\mathcal{I}^+$, it was found in \cite{Anninos:2011ui} that the number of gauge invariant degrees of freedom at $\mathcal{I}^+$ scales as $\sim N$. This is a dramatic reduction of the effective field theoretic number of degrees of freedom. Noting, further, that in higher spin theories $N \propto \tfrac{1}{G_N \Lambda}$ at large $N$, this was taken as an indication that the underlying quantum information content of the entire quantum theory should itself be of the order $\tfrac{1}{G_N \Lambda}$. The scaling behaviour $\tfrac{1}{G_N \Lambda}$ is also that expected of the (logarithm of the) Euclidean higher spin $S^4$ partition function in the semiclassical limit. Perhaps such an avenue links the Lorentzian and Euclidean picuture. 
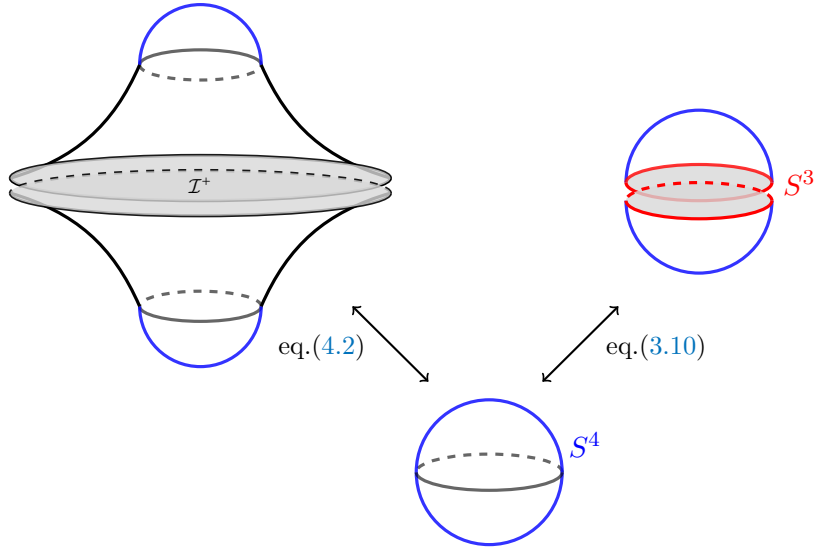
\begin{figure}[ht]
\centering
\begin{tikzpicture}

\begin{scope}
[
    scale=1,
    rotate=180,
    shift = {(-2,-5.2)},
    boundary/.style={thick},
    bulk/.style={thick, dashed},
    curve/.style={thick},
    dcurve/.style={thick, dashed},
    label/.style={scale=0.7}
]
      \draw[dashed, very thick, opacity=0.6] (1.8,1) arc [start angle=0, end angle=180, x radius=.8cm, y radius=0.2cm];
  \draw[opacity=0.8,  very thick, opacity=0.6] (0.2,1) arc [start angle=180, end angle=360, x radius=.8cm, y radius=0.2cm];
  \draw[opacity=0.8, very thick, blue] (0.2,1) arc [start angle=180, end angle=360, x radius=.8cm, y radius=.8cm];  

\coordinate (A) at (1.8,1);
\coordinate (B) at (0.2,1);

\coordinate (C) at (-1.5,2.5);
\coordinate (D) at (3.5,2.5);

\coordinate (E) at (3.5,-.5);

\draw[curve, bend right=25, very thick] (B) to (C);
\draw[curve, bend left=25, very thick] (A) to (D);

      \draw[very thick] (3.5,2.5) arc [start angle=0, end angle=180, x radius=2.5cm, y radius=0.3cm];
  \draw[opacity=0.9, very thick] (-1.5,2.5) arc [start angle=180, end angle=360, x radius=2.5cm, y radius=0.3cm];

 \fill[gray!30, opacity=0.8]
  (-1.5,2.5)
  arc[start angle=180, end angle=360, x radius=2.5cm, y radius=0.3cm]
  arc[start angle=0, end angle=180, x radius=2.5cm, y radius=0.3cm]
  -- cycle;

\end{scope}

\begin{scope}
[
    scale=1,
    boundary/.style={thick},
    bulk/.style={thick, dashed},
    curve/.style={thick},
    dcurve/.style={thick, dashed},
    label/.style={scale=0.7}
]
      \draw[dashed, very thick, opacity=0.6] (1.8,1) arc [start angle=0, end angle=180, x radius=.8cm, y radius=0.2cm];
  \draw[opacity=0.6, very thick] (0.2,1) arc [start angle=180, end angle=360, x radius=.8cm, y radius=0.2cm];
  \draw[opacity=0.8, very thick, blue] (0.2,1) arc [start angle=180, end angle=360, x radius=.8cm, y radius=.8cm];  

\coordinate (A) at (1.8,1);
\coordinate (B) at (0.2,1);

\coordinate (C) at (-1.5,2.5);
\coordinate (D) at (3.5,2.5);

\coordinate (E) at (3.5,-.5);

\draw[curve, bend right=25, very thick] (B) to (C);
\draw[curve, bend left=25, very thick] (A) to (D);

      \draw[dashed, very thick, opacity=0.9] (3.5,2.5) arc [start angle=0, end angle=180, x radius=2.5cm, y radius=0.3cm];
  \draw[opacity=0.9, very thick] (-1.5,2.5) arc [start angle=180, end angle=360, x radius=2.5cm, y radius=0.3cm];

 \fill[gray!30, opacity=0.8]
  (-1.5,2.5)
  arc[start angle=180, end angle=360, x radius=2.5cm, y radius=0.3cm]
  arc[start angle=0, end angle=180, x radius=2.5cm, y radius=0.3cm]
  -- cycle;

\node[label] at (1,2.6) {$\mathcal{I}^+$}; 
\end{scope}

\begin{scope}
[
    scale=1.2,
    rotate=180,
    shift = {(-7.3,-3.2)},
    boundary/.style={thick},
    bulk/.style={thick, dashed},
    curve/.style={thick},
    dcurve/.style={thick, dashed},
    label/.style={scale=0.7}
]

  \draw[opacity=0.8, very thick, blue] (0.2,1) arc [start angle=180, end angle=360, x radius=.8cm, y radius=.8cm];  

 \fill[gray!30, opacity=0.8]
  (1.8,1)
  arc[start angle=0, end angle=180, x radius=.8cm, y radius=0.2cm]
  arc[start angle=180, end angle=360, x radius=.8cm, y radius=0.2cm]
  -- cycle; 
 \node[red] at (-.1,1) {$S^3$}; 
      \draw[very thick,red] (1.8,1) arc [start angle=0, end angle=180, x radius=.8cm, y radius=0.2cm];
  \draw[opacity=0.8,  very thick,red] (0.2,1) arc [start angle=180, end angle=360, x radius=.8cm, y radius=0.2cm];
 
\end{scope}

\begin{scope}
[
    scale=1.2,
    shift = {(5.3,1)},
    boundary/.style={thick},
    bulk/.style={thick, dashed},
    curve/.style={thick},
    dcurve/.style={thick, dashed},
    label/.style={scale=0.7}
]

  \draw[opacity=0.8, very thick, blue] (0.2,1) arc [start angle=180, end angle=360, x radius=.8cm, y radius=.8cm];  

 \fill[gray!30, opacity=0.8]
  (1.8,1)
  arc[start angle=0, end angle=180, x radius=.8cm, y radius=0.2cm]
  arc[start angle=180, end angle=360, x radius=.8cm, y radius=0.2cm]
  -- cycle; 
  \draw[dashed, very thick,red] (1.8,1) arc [start angle=0, end angle=180, x radius=.8cm, y radius=0.2cm];
  \draw[very thick,red] (0.2,1) arc [start angle=180, end angle=360, x radius=.8cm, y radius=0.2cm];
  
\end{scope}

\begin{scope}
[
    scale=1.2,
    shift = {(3.,-2)},
    boundary/.style={thick},
    bulk/.style={thick, dashed},
    curve/.style={thick},
    dcurve/.style={thick, dashed},
    label/.style={scale=0.7}
]
  \draw[opacity=0.8, very thick, blue] (0.2,1) arc [start angle=180, end angle=360, x radius=.8cm, y radius=.8cm];  
        \draw[dashed, very thick, opacity=0.6] (1.8,1) arc [start angle=0, end angle=180, x radius=.8cm, y radius=0.2cm];
  \draw[opacity=0.8,  very thick, opacity=0.6] (0.2,1) arc [start angle=180, end angle=360, x radius=.8cm, y radius=0.2cm];
\end{scope}

\begin{scope}
[
    scale=1.2,
    rotate=180,
    shift = {(-5,0)},
    boundary/.style={thick},
    bulk/.style={thick, dashed},
    curve/.style={thick},
    dcurve/.style={thick, dashed},
    label/.style={scale=0.7}
]
  \draw[opacity=0.8, very thick, blue] (0.2,1) arc [start angle=180, end angle=360, x radius=.8cm, y radius=.8cm];  
\node[blue] at (-.05,.7) {$S^4$};
  
\end{scope}

\draw[<->,thick] (4,0)-- (3,1);
\draw[<->,thick] (5.5,0)-- (6.5,1);
\node[scale=.9] at (2.6,.45) {eq.(\ref{eq: Norm HH})};
\node[scale=.9] at (7,.45) {eq.(\ref{eq:HS_S4})};

\end{tikzpicture}
\caption{Interplay between the Euclidean $S^4$ partition function and the Lorentzian Hartle-Hawking wavefunction of a $\Lambda>0$ higher spin theory. The one-loop analysis motivates the gluing formula (\ref{eq:HS_S4}), which suggests that the $S^4$ partition function is obtained by gluing two hemispheres with a common  three-sphere boundary. On the boundary we place the $\mathrm{Sp}(N)$ invariant sector of a $\mathrm{Sp}(N)$ vector model, coupled to conformal higher spin sources. Strikingly, the $\mathrm{Sp}(N)$ theory which appears in this bilinear pairing also encodes the Lorentzian Hartle-Hawking wavefunction computed in the context of higher spin dS$_4$/CFT$_3$. This suggests that the sphere partition function captures aspects of the wavefunction norm (\ref{eq: Norm HH}).}
\label{fig:HH}
\end{figure}
\newline\newline
The question we address in this paper is the following: What is the microphysical completion of a $\Lambda>0$ higher spin theory on $S^4$? To attack this problem, we will consider the contribution to the sphere path integral from the full higher spin spectrum at one-loop \cite{Anninos:2020hfj}. Ordinarily, little can be hoped for from a one-loop calculation. But when we have more complete knowledge of the perturbative particle spectrum, properties of the more complete theory might be anticipated. For instance, an analogous one-loop analysis for certain Euclidean AdS$_4$ higher spin theories anticipates the duality \cite{Klebanov:2002ja} to a free $O(N)$ model \cite{Giombi:2013fka}. In perturbative string theory, summing over the one-loop contributions of the target space particle content reveals the structure of a genus one worldsheet path integral. Thus, in fortuitous circumstances,  we may hope to reveal more structure than what we bargained for, even at the one-loop level. We will see glimmers of such fortuity.
\newline\newline
The paper is organised as follows. In section \ref{sec2}, we briefly review some properties of Fronsdal's higher spin gauge fields. We comment on their representation theoretic content, and their individual one-loop contributions to the sphere path integral. In section \ref{sec3}, we consider higher spin spectra with bosonic field content alone, and analyse their one-loop sums. For the sum over only the even spins, known as the minimal spectrum, the result metamorphoses into two terms, (\ref{eq:ZHS}) and (\ref{eq:Zfree}). The first is identified as the one-loop contribution of a conformal higher spin gauge theory in three-dimensions, while the second is found to be {{twice}} the partition function of a free conformally coupled scalar on $S^3$. We interpret this result as the first subleading term of the large $N$ expansion of a path integral (\ref{eq:HS_S4}) over conformal higher spin gauge fields interacting with two copies of a  free $\text{Sp}(N)$ conformal field theory. Whether a gluing formula of the type (\ref{eq:HS_S4}), reminiscent of expressions in topological quantum field theory, is a defining property of $\Lambda>0$ quantum gravity is left as an interesting question. Although our calculations are in a Euclidean setting, we find notable features reminiscent of a type of dS$_4$/CFT$_3$ picture. This is explored in section \ref{sec4}, and summarised in figure \ref{fig:HH}. In section \ref{sec5}, we consider higher spin models with fermionic and bosonic fields \cite{Sezgin:2012ag,Chang:2012kt,Hertog:2017ymy}. Here, the corresponding completion is proposed to be an $\mathcal{N}=2$ extension of (\ref{eq:HS_S4}). Due to a series of remarkable cancellations, the leading order result is $\mathcal{Z}[S^4] \approx 2^N$, with $N\in\mathbb{Z}^+$, and we further observe complete one-loop cancellations up to a universal group volume contribution. It is conceivable that the $\mathcal{N}=2$ partition function is amenable to supersymmetric localisation methods and hence one-loop exact. This result opens up an interesting quantitative avenue towards a microscopical interpretation of the de Sitter horizon entropy. We end with an outlook in section \ref{sec6}, emphasising technical and conceptual questions.  

\section{$\Lambda>0$ higher spin theory}\label{sec2}

In this section, we discuss the basic field content of the non-supersymmetric higher spin theories in dS$_4$. This consists of massless fields of arbitrary spin plus a conformally coupled scalar.  

\subsection{Fronsdal fields}

A free totally massless spin-$s$ field in dS$_4$, with $s\in \mathbb{Z}^+$, is a totally symmetric tensor valued field $b_{\nu_1 \nu_2 \ldots \nu_s}$ which satisfies
\begin{eqnarray}\nonumber
   0&=&b^{\nu_1 \nu_2}_{\quad ~ \nu_1 \nu_2 \mu_3 \ldots \mu_{s-2}}~,\\ \nonumber
    0&=&\nabla_\nu \nabla^\nu b_{\mu_1 \ldots \mu_s} - s\nabla_\nu \nabla_{(\mu_1} b^\nu_{~~\mu_2 \ldots \mu_s)} + \frac{1}{2}s(s-1) \nabla_{(\mu_1} \nabla_{\mu_2} b^\nu_{~~\nu \mu_3 \ldots \mu_s)} -2 (s-1)(s+1)b_{\mu_1\ldots \mu_s}~,
 \end{eqnarray}   
where the double-traceless conditions on $b_{\nu_1 \nu_2 \ldots \nu_s}$ appear at $s\ge 4$. We have set the de Sitter length $\ell_{\text{dS}}=1$.   The equations of motion are known as the Fronsdal equations \cite{Fronsdal:1978rb} and are invariant under
\begin{equation}
    b_{\mu_1 \ldots \mu_s} \rightarrow b_{\mu_1 \ldots \mu_s} + \nabla_{(\mu_1}\xi_{\mu_2 \ldots \mu_s)}~,\quad\quad {\xi}^\nu_{~~\nu\mu_3 \ldots \mu_{s-3}}=0~,
\end{equation}
where the traceless conditions on the gauge-parameter $\xi_{\mu_2 \ldots \mu_s}$ appear at $s\ge 3$. Each Fronsdal field carries two locally propagating degrees of freedom. The Fronsdal fields play a crucial role in higher spin models of de Sitter space \cite{Vasiliev:1990en,Vasiliev:2003ev,Iazeolla:2007wt} --- they constitute the entire perturbative field content. They can also be defined on a four-sphere background in an equivalent way. We denote the set of spin $s \geq 0$ Fronsdal fields by $\{b, b_{\nu_1}, b_{\nu_1\nu_2}, b_{\nu_1\nu_2\nu_3}, \ldots\}$.

\subsection{Half-integer higher spin fields}

There are also half-integer higher spin fields in four-dimensional de Sitter space \cite{Fang:1978wz, Fang:1979hq}. These are less studied than their bosonic counterpart, however they also obey simple wave equations. For spin-$\tfrac{3}{2}$ one has the Rarita-Schwinger gauge field equation \cite{Fierz:1939ix, Rarita:1941mf,Fang:1979hq, Letsios:2022tsq}:
\begin{equation}
\gamma^{\mu\nu\rho}\left(  \nabla_\nu + \frac{\ii}{2} \gamma_\nu \right) \Psi_\rho = 0~, \quad\quad \Psi_\mu \to \Psi_\mu + \left(\nabla_\mu +\frac{\ii}{2}\gamma_\mu\right)\lambda ~,
\end{equation}
where we now  have a spinor valued gauge parameter $\lambda$. The Dirac Rarita-Schwinger field  $\Psi_\mu$ encodes four real propagating degrees of freedom. One can similarly obtain the wave-equations for gauge fields of arbitrary half-integer spin \cite{Deser:2003gw,Letsios:2023awz}. Their Dirac fields also encode four real propagating degrees of freedom.

\subsection{de Sitter representation theory} 

From a group theoretical perspective, one can classify free fields in de Sitter space according to the unitary irreducible representations of the dS$_4$ isometry group $\mathrm{SO}(1,4)$, or its double cover $\mathrm{Spin}(1,4)$ if we want to include fermionic representations. In either case the Lie algebra is $\mathfrak{so}(1,4)$. We summarize the bosonic representations in table \ref{tab:scalar-reps-dS4}.

\begin{table}[ht]
    \centering
    \small
         \setlength{\arrayrulewidth}{1.2pt}
    \renewcommand{\arraystretch}{1.5}
    \begin{tabular}{|c!{\vrule width .8pt}c!{\vrule width .8pt}c|}
        \hline
        Irrep & Range of $\Delta$ & Range of $s$ \\
        \hhline{|=|=|=|}
        $\pi_\nu$ 
        & $\Delta = \tfrac{3}{2} + \ii\nu,\ \nu \in \mathbb{R}$ 
        & $s = 0,1,2,\ldots$ \\
        \hhline{|=|=|=|}
        \multirow{2}{*}{$\gamma_\Delta$}
        & $0 < \Delta < 3$
        & $s = 0$ \\
        \hhline{|~|~|~|}
        & $0 < \Delta < 2$
        & $s \geq 1$ \\
        \hhline{|=|=|=|}
        $\mathcal{E}_{\Delta,0}$ 
        & $\Delta = 2 + p,\ p \geq 1$ 
        & $s = 0$ \\
        \hline
        $D_{s,t}^{\pm}$ 
        & $\Delta = 2 + t,\ t = 0,1,\ldots,s-1$ 
        & $s \geq 1$ \\
        \hline
    \end{tabular}
    \caption{Summary of bosonic representations of $\mathfrak{so}(1,4)$. }
    \label{tab:scalar-reps-dS4}
\end{table}

\begin{table}[ht]
    \centering
    \small
         \setlength{\arrayrulewidth}{1.2pt}
    \renewcommand{\arraystretch}{1.5}
    \begin{tabular}{|c!{\vrule width .8pt}c!{\vrule width .8pt}c|}
        \hline
        Irrep & Range of $\Delta$ & Range of $s$ \\
        \hhline{|=|=|=|}
        $\pi_{\nu}$ 
        & $\Delta = \tfrac{3}{2}+\ii\nu,~\nu \in \mathbb{R}^+$ 
        & $s=\frac{1}{2},\frac{3}{2},\frac{5}{2},\ldots$ \\
        \hhline{|=|=|=|}
        $\gamma_\Delta$ 
        & $\mathsf{X}$ 
        &  $\mathsf{X}$  \\
        \hhline{|=|=|=|}
        $D_{s,t}^\pm$ 
        & $\Delta = 2+t~,~t= -\frac{1}{2},\frac{1}{2},\ldots, s-1$
        &  $s=\frac{1}{2},\frac{3}{2},\ldots$   \\
        \hline
    \end{tabular}
    \caption{Summary of fermionic representations of $\mathfrak{so}(1,4)$. In particular there are no half-integer complementary series irreducible representations in $\mathfrak{so}(1,4)$.}
    \label{tab:fer-reps-dS4}
\end{table}

\noindent For the bosonic case we have three types of unitary irreducible representations: principal series $\pi_\nu$, complementary series $\gamma_\Delta$, spinless exceptional  series $\mathcal{E}_{\Delta,0}$, and discrete series $D_{s,t}^\pm$. The principal series describes heavy fields in de Sitter space. The complementary series irrep contains light fields. Finally, gauge fields transform in the discrete series irrep. Fields in the discrete series $D_{s,t}^\pm$ are called partially massless fields (PMF) of depth $t$. 
Group theoretically the $\pm$ superscript comes from the fact that $\mathrm{SO}(4) \cong \mathrm{SU}(2)\times \mathrm{SU}(2)$ \cite{Higuchi:1991tn, RiosFukelman:2023mgq, Letsios:2022tsq, Letsios:2023qzq}.  
Except for the scalar $b$ which transforms in the complementary series with $\Delta_b=1$, the Fronsdal fields (\ref{eq:B}) $b_{\nu_1 \ldots\nu_s}$ are spin $s$ highest depth $t=s-1$ partially massless fields with weight $\Delta= 1+s$. 
\newline\newline
Physically, the highest depth PMF correspond to `massless' gauge fields: the photon at spin one, the graviton at spin two, and so on. 
The single-particle Hilbert space of highest depth PMF furnishes a  discrete series unitary irreducible representation of $\mathrm{SO}(1,4)$  denoted by $D_{s,s-1}^{\pm}$. 
In general, PMF are fields below the Higuchi bound \cite{Higuchi:1986py,Higuchi:1986wu,Higuchi:1989gz}. Whereas in general fields below the Higuchi bound are non-unitary and have $2s+1$ degrees of freedom, at specific points gauge redundancy removes the ghost-like non-unitary degrees of freedom, and yields the partially massless unitary fields. This comes at the price of reducing the number of degrees of freedom: highest depth PMF only have two degrees of freedom (akin to the two helicities).

\subsection{One-loop higher spin sphere partition function}

In order to discuss the Euclidean sphere partition function of a $\Lambda>0$ higher spin theory we need one more ingredient. Schematically, this ingredient is:
\begin{equation}\label{eq:relation}
\begin{array}{@{}c@{\hspace{2em}}c@{\hspace{2em}}c@{}}
\text{\shortstack[c]{Lorentzian Harish-Chandra \\ character}} 
& \raisebox{0.5ex}{$\displaystyle \Longleftrightarrow$} 
& \text{\shortstack[c]{Euclidean one-loop sphere \\ partition function}}
\end{array}
\end{equation}
Given a unitary irreducible representation, one can write down the corresponding Harish-Chandra character. The Harish-Chandra character is a generalization of the group character of compact groups to non-compact groups such as $\mathrm{SO}(1,4)$. In particular, for a representation $\mathcal{R}$ of $\mathfrak{so}(1,4)$ classified by its weight and spin, $(\Delta,s)$, we have 
\begin{equation}
    \chi_{\Delta,s}(\mathfrak{t}) = \tr_{\Delta,s} \e^{-\ii \mathfrak{t} H}~,
\end{equation}
where the trace runs over all the states in $\mathcal{R}$. We restrict to $H \in \mathfrak{so}(1,1)$, which generates time translations in the static patch. We note that de Sitter space does not admit everywhere timelike Killing vector fields, but they do exist if we restrict to a single static patch.
\newline\newline
The Harish-Chandra characters are intrinsically Lorentzian objects associated with representations of the de Sitter isometry group. Nevertheless, the right-hand side of \eqref{eq:relation} establishes a nontrivial connection between these Lorentzian group characters and partition functions computed on a spherical topology. The sphere, in turn, is the geometry of Euclidean de Sitter space.
As shown in \cite{Anninos:2020hfj}, the one-loop partition function on the sphere can be expressed directly in terms of Lorentzian characters. For example, a conformally coupled scalar on a unit $S^4$ with $m^2\ell_{\mathrm{dS}}^2 =2$ satisfies
\begin{equation}\label{eq:PI_characters}
    \log \mathcal{Z}_{1,0}^{(1)} = \log {\det}^{-\frac{1}{2}}\Big(\frac{-\nabla^2 +2 \ell_{\mathrm{dS}}^{-2}}{\Lambda_{\text{u.v.}}}\Big) = \int_0^\infty \frac{\d\mathfrak{t}}{2\mathfrak{t}} \frac{1+\e^{-\mathfrak{t}}}{1-\e^{-\mathfrak{t}}} \frac{\e^{-\mathfrak{t}} + \e^{-2\mathfrak{t}}}{(1-\e^{-\mathfrak{t}})^3}~,
\end{equation}
where $-\nabla^2$ is the Laplacian on $S^4$, and $\Lambda_{\text{u.v.}} = {\ell_{\mathrm{u.v.}}^{-2}}$ is some ultraviolet cutoff scale with units of inverse length squared, which has been formally removed in the final equality. The integrand is the Harish-Chandra character of a scalar with $\Delta_+=2$ (or equivalently $\Delta_-=1$) in the complementary series irreducible representation of $\mathrm{SO}(1,4)$. 
\newline\newline
As a general remark, we note that the final equality in (\ref{eq:PI_characters}) (as well as similar integrals used throughout the text) should be understood as a formal expression that more correctly depends, locally, on the ultraviolet cutoff. In four dimensions, one will generally have three types of divergences, namely $\sim\Lambda_{\text{u.v.}}^{2}$, $\sim\Lambda_{\text{u.v.}}^{}$, and $\sim\log \Lambda_{\text{u.v.}}$. These are associated to a renormalisation of the cosmological constant $\Lambda$, the Newton constant $G_N$, and the dimensionless coupling $\vartheta$ of the Gauss-Bonnet higher derivative term. Due to the logarithmic divergence, the overall prefactor of the four-sphere path integral is not a priori fixed, or stated otherwise, one must fix the physical coupling $\vartheta$. On the other hand, for a three-dimensional parity invariant theory, one only has $\sim\Lambda_{\text{u.v.}}^{3/2}$, and $\sim\Lambda_{\text{u.v.}}^{1/2}$, local divergences. There is no room to renormalise the constant prefactor of the three-sphere path integral due to the absence of dimensionless couplings in the gravitational effective field theory. As such, the three-sphere path integral is less ambiguous than the four-sphere path integral. Having said that, we should also note that in theories with additional structure, not all counterterms are allowed and this may render the four-dimensional path integral less ambiguous as well.\footnote{For instance though the Gauss-Bonnet higher derivative term is permitted in a generic gravitational effective field theory, it may not be allowed in a higher spin gravity theory where it would have to be extended to a full higher spin gauge invariant term. Interestingly, as we shall see in what follows, adding up all the four-dimensional one-loop divergences of higher spin fields yields divergences that more closely resemble those of a three-dimensional) rather than four-dimensional theory. As such, there is no logarithmic divergence in the resulting sum, leaving less room for ambiguities in the overall constant pre-factor. Furthermore, one can always consider judiciously chosen ratios of partition functions.}
\newline\newline
The general form of (\ref{eq:PI_characters}) remains the same for arbitrary $(\Delta,s)$. For partially massless fields the one-loop sphere partition function also encodes a codimension-two character in addition to the  $\mathrm{SO}(1,4)$ `bulk' Harish-Chandra character. Concretely, the formal expression  for the totally massless case is
\begin{equation}
    \log \mathcal{Z}_{1+s,s}^{(1)} = \int_0^\infty \frac{\d\mathfrak{t}}{2\mathfrak{t}} \frac{1+\e^{-\mathfrak{t}}}{1-\e^{-\mathfrak{t}}}\chi_{1+s, s}(\mathfrak{t}) ~,
\end{equation}
where $\chi_{1+s, s}(\mathfrak{t}) \equiv \chi_{s,\mathrm{bulk}}(\mathfrak{t}) - \chi_{s,\mathrm{edge}}(\mathfrak{t})$ is the difference between a SO(1,4) Harish-Chandra character (the bulk part) and the edge contribution which we denote as $\chi_{s,\mathrm{edge}}(\mathfrak{t})$. Generally speaking the above expression will suffer from ultraviolet divergences which can be treated with the addition of local counteterms.
\newline\newline
The one-loop contributions for the higher spin fields $\{b, b_{\nu_1}, b_{\nu_1\nu_2}, b_{\nu_1\nu_2\nu_3}, \ldots\}$ are \cite{Anninos:2020hfj}
\begin{subequations}\label{eq:chis}
\begin{align}\label{eq:chi10}
\chi_{1, 0}(\mathfrak{t}) &= \frac{\e^{-\mathfrak{t}} + \e^{-2\mathfrak{t}}}{(1-\e^{-\mathfrak{t}})^3}~,\\ \label{eq:chissp1}
\chi_{1+s, s}(\mathfrak{t}) &= \chi_{s,\mathrm{bulk}}(\mathfrak{t}) - \chi_{s,\mathrm{edge}}(\mathfrak{t}) ~,\cr
     \chi_{s,\mathrm{bulk}}(\mathfrak{t})&= 2\frac{(2s+1)\e^{-(s+1)\mathfrak{t}} -(2s-1)\e^{-(s+2)\mathfrak{t}}}{(1-\e^{-\mathfrak{t}})^3}~,\\ \label{eq:chiedge}
    \chi_{s,\mathrm{edge}} (\mathfrak{t})&=\frac{1}{3} \frac{s(s+1)(2s+1) \e^{-s \mathfrak{t}} -s(s-1)(2s-1) \e^{-(s+1)\mathfrak{t}}}{(1-\e^{-\mathfrak{t}})}~.
 \end{align}   
\end{subequations}
The terminology of `bulk' versus `edge' refers to the degree of divergence exhibited by the characters in the small $\mathfrak{t}$ limit. For the bulk character of dS$_{1+3}$ the divergence goes as $\tfrac{1}{\mathfrak{t}^3}$. Instead, the edge character (\ref{eq:chiedge}) exhibits a codimension-two divergence of the form $\tfrac{1}{\mathfrak{t}}$.

\section{$S^4$ partition function}\label{sec3}

We will now focus on the Euclidean path integral of  higher spin theories around the four-sphere saddle. 
More specifically, we will consider the particle spectrum of {minimal} and {non-minimal} higher spin theories which differ by their field content. The minimal higher spin spectrum features a real scalar field, $b$, and even spin-$s$ fields, $b_{\nu_1 \nu_2\ldots \nu_s}$. The non-minimal higher spin spectrum features a real scalar, and both even and odd higher spin fields. That is:
\begin{subequations}\label{eq:B}
    \begin{align}
        \text{Minimal higher spin:}&\quad  \{b, b_{\nu_1\nu_2}, b_{\nu_1\nu_2\nu_3\nu_4}, \ldots\}~,\\
        \text{Non-minimal higher spin:}&\quad  \{b, b_{\nu_1}, b_{\nu_1\nu_2}, b_{\nu_1\nu_2\nu_3}, \ldots\}~.
    \end{align}
\end{subequations}
In both cases, the scalar field, $b$, is a conformally coupled scalar of mass $m^2\ell_{\text{dS}}^2 = +2$. The minimal and non-minimal theories can be further  categorised into two types, type A and type B, which have the same spectra but different parity structure. The scalar field in the type B theory, for instance, is a pseudo-scalar. We will focus on the minimal type A theory. 
\newline\newline
At one-loop order we use the relation (\ref{eq:relation}), which allows us to write down the one-loop sphere partition function without knowledge of the higher spin action. One must sum over all the perturbative particle content. This sum yields the (formal) expression
\begin{equation}\label{oneloophs}
     \log \mathcal{Z}_{\mathrm{h.s.}}^{(1)} =\int_0^\infty\! \frac{\d \mathfrak{t}}{2\mathfrak{t}} \frac{1+\e^{-\mathfrak{t}}}{1-\e^{-\mathfrak{t}}}  \sum_{s\in 2\mathbb{N}}\chi_{1+s,s}(\mathfrak{t}) 
     = \int_0^\infty \frac{\d \mathfrak{t}}{2\mathfrak{t}} \frac{1+\e^{-\mathfrak{t}}}{1-\e^{-\mathfrak{t}}} \bigg(\frac{-\e^{-\mathfrak{t}}}{(1-\e^{-\mathfrak{t}})^2} +2 \frac{\e^{-\frac{3\mathfrak{t}}{2}}+\e^{-\frac{\mathfrak{t}}{2}}}{(1-\e^{-\mathfrak{t}})^2}\bigg)~,
\end{equation}
where in going to the second equality we performed the transformation $\mathfrak{t}\rightarrow \tfrac{\mathfrak{t}}{2}$. 
The two pieces on the right hand side are identified as follows:
\begin{enumerate}
    \item The first part is
    \begin{equation}\label{eq:ZHS}
        \log \mathcal{Z}_{\mathrm{HS}} =  \int_0^\infty \frac{\d \mathfrak{t}}{2\mathfrak{t}}\frac{1+\e^{-\mathfrak{t}}}{1-\e^{-\mathfrak{t}}}\frac{-\e^{-\mathfrak{t}}}{(1-\e^{-\mathfrak{t}})^2}~.
    \end{equation}
    This is the one-loop partition function of a three-dimensional conformal higher spin gauge theory on $S^3$ (see Appendix \ref{app:fermionic}). 
    \item The second part is 
    \begin{equation}\label{eq:Zfree}
        \log \mathcal{Z}_{\mathrm{free}} = 2\int_0^\infty \frac{\d \mathfrak{t}}{2\mathfrak{t}}\frac{1+\e^{-\mathfrak{t}}}{1-\e^{-\mathfrak{t}}}\frac{\e^{-\frac{3\mathfrak{t}}{2}}+\e^{-\frac{\mathfrak{t}}{2}}}{(1-\e^{-\mathfrak{t}})^2}~.
    \end{equation}
    This is twice the partition function of a free conformally coupled real scalar on $S^3$ with mass $m^2 \ell_{\mathrm{dS}}^2 =\tfrac{3}{4}$. Viewed as a scalar on Euclidean dS$_3$, it transforms in the complementary series irreducible representation of the dS$_3$ isometry group $\mathrm{SO}(1,3)$ with $\Delta_+= \frac{3}{2}$ (or equivalently $\Delta_- = \frac{1}{2}$).
\end{enumerate}
Employing a heat-kernel regularisation scheme \cite{Klebanov:2011gs,Anninos:2020hfj}, one finds the regularised expression
\begin{equation}\label{eq:evaluated_ccscalar}
   \int_\varepsilon \frac{\d \mathfrak{t}}{2\mathfrak{t}}\frac{1+\e^{-\mathfrak{t}}}{1-\e^{-\mathfrak{t}}}\frac{\e^{-\frac{3\mathfrak{t}}{2}}+\e^{-\frac{\mathfrak{t}}{2}}}{(1-\e^{-\mathfrak{t}})^2}=  - \frac{1}{8}\log 2 + \frac{3\zeta(3)}{16\pi^2} + \frac{\pi}{2\varepsilon^3} +\frac{\pi}{16\varepsilon}~.
\end{equation}
The parameter $\varepsilon$ is related to the ultraviolet cutoff as $\varepsilon \sim \Lambda_{\mathrm{u.v.}}^{-1/2}$, and the divergent terms can be reabsorbed into  local counterterms. What cannot be removed with local counterterms is the real-valued constant prefactor $- \tfrac{1}{8}\log 2 + \tfrac{3\zeta(3)}{16\pi^2}$. Similarly, we obtain 
\begin{equation}\label{eq:evalated_HS}
      \int_\varepsilon \frac{\d \mathfrak{t}}{2\mathfrak{t}}\frac{1+\e^{-\mathfrak{t}}}{1-\e^{-\mathfrak{t}}}\frac{-\e^{-\mathfrak{t}}}{(1-\e^{-\mathfrak{t}})^2} = +\frac{\zeta(3)}{8\pi^2} + \mathcal{O}({\varepsilon^{-3},\varepsilon^{-1}})~.
\end{equation}
The integral was evaluated using a zeta-function regularisation scheme in \cite{Klebanov:2011gs,Giombi:2013yva}. The same result follows from a heat-kernel analysis, where the divergent terms can be removed with three-dimensional local counterterms. 
\newline\newline
The result \eqref{oneloophs} for the one-loop contribution is rather remarkable. 
On the right-hand side of~\eqref{oneloophs}, we compute the four-sphere partition 
function of a higher spin theory with positive cosmological constant, 
$\Lambda>0$. This involves summing over all four-dimensional characters~\eqref{eq:chis} 
for spins $s \in 2\mathbb{N}_0$. In general dimensions, such a sum cannot be performed 
explicitly; indeed, this resummation appears to be possible only for dS$_4$ and dS$_2$~\cite{Beatrix_Toda}.
\newline\newline
Strikingly, the second line of~\eqref{oneloophs} --- corresponding to the sum over all four-dimensional higher spin characters --- organizes itself into a three-dimensional character. As a result, it captures the structure of a three-sphere partition function, including its characteristic divergences. This raises an immediate and  conceptually sharp question: since the four-sphere is a closed four-manifold, how should we interpret the appearance of three-dimensional divergences?
One can reasonably expect that this feature provides an important clue about the structure of the dual theory, as has  been observed in the higher spin AdS$_4$ context~\cite{Giombi:2013fka}. In the 
following section, we will propose a possible completion of the one-loop 
contribution~\eqref{oneloophs} that addresses this puzzle.
\newline\newline
Before we proceed, we combine the one-loop higher spin contribution with the higher spin on-shell action. 
To leading order, one would have minus the on-shell action, $-S_{\text{dS}_4}$, of the higher spin theory evaluated on $S^4$. While the value of this on-shell action is not known, certain things about it can be anticipated. For one, we expect the on-shell action to scale as $\tfrac{1}{G_N\Lambda}$, where $G_N$ is the four-dimensional Newton constant. Secondly, based on observations in Einstein gravity, the leading order (finite part) of the Euclidean AdS$_4$ action is minus one-half that of Euclidean dS$_4$ (or equivalently, $S^4$). 
\newline\newline
Implementing the AdS$_4$/CFT$_3$ correspondence \cite{Klebanov:2002ja} relating higher spin theory on Euclidean AdS$_4$ to the free $\mathrm{O}(N)$ vector model, we can postulate a value of $S_{\text{dS}_4}$. We note that $\frac{1}{G_N \Lambda} \propto N$, to leading order in the large $N$ limit, but the relation between the two can have subleading contributions. For a more general vector model where the rank goes as $N$, one might anticipate that $\frac{1}{G_N \Lambda} = \alpha N + \beta$, for some real $N$-independent constants $\alpha$ and $\beta$. The on-shell action of Euclidean AdS$_4$, $S_{\text{AdS}_4}^{(N)}$, can be deduced from the total one-loop contribution of the bulk fields calculated in \cite{Giombi:2013fka}, combined with the $S^3$ partition function of the free $O(N)$ model computed for instance in \cite{Klebanov:2011gs}.\footnote{In fact, this is somewhat imprecise, since upon gauging the $O(N)$ group we should generally include an $O(N)$ Chern-Simons gauge theory at level $k$. Even at parameterically small coupling, $k\to\infty$, the $S^3$ partition function of the decoupled Chern-Simons theory will contribute an overall factor. These factors are suppressed in our expressions throughout. It is an interesting open question what the bulk AdS$_4$ dual to the decoupled Chern-Simons sector is, which plays an increasingly interesting role as one increases the complexity of the boundary topology \cite{Banerjee:2012gh}. The Chern-Simons contribution will scale as $\sim -\frac{N^2}{4}\log k$ at large $N$ in the leading weak coupling limit, as opposed to the $\sim N$ scaling of the matter sector. It is less clear, but an interesting possibility, that such an additional decoupled bulk sector should also be present in the de Sitter theory at trivial topology.}  We then postulate the on-shell actions of the higher spin Euclidean AdS$_4$ and $S^4$ to be related as $S_{\text{dS}_4}^{(N)} = 2 S_{\text{AdS}_4}^{(-N)}$. This can be viewed as a refined version of the naive analytic continuation $\Lambda\to-\Lambda$ in terms of an unambiguous parameter, $N$, of the underlying theory. We then find
\begin{equation}\label{eq:S4}
-S_{\text{dS}_4} = 2(N+1)  \left( \frac{1}{8}\log 2 - \frac{3\zeta(3)}{16\pi^2} \right) =    (N+1) \times  0.1276 \ldots~.
\end{equation}
We note that (\ref{eq:S4}), which according to Gibbons and Hawking \cite{Gibbons:1977mu, Gibbons:1976ue} computes the leading contribution to the de Sitter horizon entropy, is positive much like it is for ordinary Einstein gravity with $\Lambda>0$. 
\newline\newline
\noindent 
Upon incorporating the one-loop contribution (\ref{oneloophs}) to the on-shell action (\ref{eq:S4}), we obtain (up to local divergences)
\begin{equation}\label{oneloopN_4D}
    \mathcal{Z}_{\mathrm{h.s.}}^{(N)}[S^4]\approx \frac{(-\ii)^{\mathcal{P}}}{\mathrm{vol}_N\,\mathcal{G}_{\mathrm{HS}}}\times \e^{-S_{\mathrm{dS}_4}} \times  \mathcal{Z}_{\mathrm{h.s.}}^{(1)}= \frac{(-\ii)^{\mathcal{P}}}{\mathrm{vol}_N\,\mathcal{G}_{\mathrm{HS}}}\times \e^{2{N} \left( \frac{1}{8}\log 2 - \frac{3\zeta(3)}{16\pi^2} \right) +\frac{\zeta(3)}{8\pi^2}}~,
\end{equation}
where $\mathcal{Z}_{\mathrm{h.s.}}^{(1)}$ is given in (\ref{oneloophs}). We have denoted by $\mathcal{G}_{\mathrm{HS}}$ the higher spin group, and by $\mathrm{vol}_N\,\mathcal{G}_{\mathrm{HS}}$ its volume, which can a priori depend on $N$. The reason for this dependence is that, as computed by the path integral, the generators of the group are normalised with respect to the coupling constant (see appendix G of \cite{Anninos:2020hfj}). We note that $\mathcal{G}_{\mathrm{HS}}$ can be represented in either a four-dimensional `bulk' way as acting on the various Fronsdal fields \cite{Vasiliev:2003ev} or a three-dimensional `boundary' way in terms of the symmetries of the conformal Laplace operator \cite{Eastwood:2002su}. To be more precise, by $\mathcal{G}_{\mathrm{HS}}$ we refer to a particular real form of the complexification of the standard higher spin group which ordinarily admits either $\mathrm{SO}(1,4)$ or $\mathrm{SO}(2,3)$ as a subgroup. The real form we are interested in is a higher spin group that admits an $\mathrm{SO}(5)$ subgroup, as explored in \cite{Iazeolla:2007wt}, which is the isometry group of the round four-sphere. The higher spin group of interest is generated by the  set of symmetric Killing tensor fields on $S^4$ at arbitrary rank.
\newline\newline
The quantity $\mathcal{P}$ encodes the generalization of Polchinski's phase \cite{Polchinski:1988ua} for the one-loop path integral on the sphere. On the four-sphere it is found to be  \cite{Anninos:2020hfj}
\begin{equation}\label{totP}
\mathcal{P} \equiv \sum_{s\in 2\mathbb{N}_0} P_s~, \quad\quad \text{where} \quad\quad    P_s= \frac{s(s^2-1)^2}{3}~.
\end{equation}
For even spin $s\in 2\mathbb{N}_0$ we have $(\pm \ii)^{P_{s}} =(-1)^\frac{s}{2}$, such that $(\pm \ii)^{P_{0}}=+1$, $(\pm \ii)^{P_{2}}=-1$, $(\pm \ii)^{P_{4}}=+1$, $(\pm \ii)^{P_{6}}=-1$, and so on. As $\mathcal{P}$ depends on an infinite non-convergent sum, it is unclear whether we can assign a definite value to $\mathcal{P}$. Recently, a proposal for a regularization of $\mathcal{P}$ was made in \cite{Giombi:2026sqa}, who argue $\mathcal{P}=-\tfrac{1}{8}$ for the minimal model and $\mathcal{P}=0$ for the non-minimal model.

\subsection{A gluing formula}
\begin{figure}[ht]
\begin{center}
\begin{tikzpicture}[scale=.7]
\begin{scope}
  \shade[ball color=gray!10!gray, opacity=0.35] (0,1) circle (2cm);
  \draw[thick] (0,1) circle (2cm);
  \draw[dashed, very thick, opacity=0.6, red] (2,1) arc [start angle=0, end angle=180, x radius=2cm, y radius=0.5cm];
  \draw[opacity=0.8, very thick, red] (-2,1) arc [start angle=180, end angle=360, x radius=2cm, y radius=0.5cm];
\shade[top color=gray!10, bottom color=gray!80, shading angle=90, opacity=0.6]
  (6,0)
  arc[start angle=180, end angle=360, x radius=2cm, y radius=2cm]   
  -- (10,0.)                                                       
  arc[start angle=360, end angle=180, x radius=2cm, y radius=0.5cm] 
  -- cycle;  
\shade[top color=gray!10, bottom color=gray!80, shading angle=90, opacity=0.6]
  (10,2) arc [start angle=0, end angle=180, x radius=2cm, y radius=0.5cm]  
  arc [start angle=180, end angle=360, x radius=2cm, y radius=-2cm]        
  -- cycle;
\fill[gray!20, opacity=0.5]
  (6,0) 
  arc [start angle=180, end angle=360, x radius=2cm, y radius=0.5cm] 
  -- (10,0)                                                        
  arc [start angle=0, end angle=180, x radius=2cm, y radius=0.5cm] 
  -- cycle; 
\fill[gray!20, opacity=0.9]
  (6,2)
  arc [start angle=180, end angle=360, x radius=2cm, y radius=0.5cm]  
  arc [start angle=0, end angle=180, x radius=2cm, y radius=0.5cm]    
  -- cycle; 
 \draw[opacity=0.8, thick, black] (6,0) arc [start angle=180, end angle=360, x radius=2cm, y radius=2cm];   
\draw[opacity=0.8, thick, black] (10,2) arc [start angle=360, end angle=180, x radius=2cm, y radius=-2cm];
 \draw[opacity=0.8, very thick, red] (6,0) arc [start angle=180, end angle=360, x radius=2cm, y radius=0.5cm];
 \draw[very thick, opacity=0.8, red] (10,0) arc [start angle=0, end angle=180, x radius=2cm, y radius=0.5cm];
\draw[opacity=0.8,very thick, red] (6,2) arc [start angle=180, end angle=360, x radius=2cm, y radius=0.5cm]; 
 \draw[very thick, opacity=0.8, red] (10,2) arc [start angle=0, end angle=180, x radius=2cm, y radius=0.5cm];
\draw[very thick, blue, decorate,
      decoration={snake, amplitude=0.07cm, segment length=0.5cm}] (8,0) -- (8,2); 
\draw[very thick, blue, decorate,
      decoration={snake, amplitude=0.07cm, segment length=0.5cm}] (7,0) -- (7,2);       
\draw[very thick, blue, decorate,
      decoration={snake, amplitude=0.07cm, segment length=0.5cm}] (9,0) -- (9,2);          
\node[scale=1, red] at (10.5,2.1) {$S^3$}; 
\node[scale=1, red] at (10.5,0) {$S^3$};     

\node[scale=1] at (.1,3.35) {$S^4$};  

\end{scope}
\end{tikzpicture}
\caption{The sphere partition function of four-dimensional higher spin theory with $\Lambda>0$ can be obtained by gluing together two hemispheres with an $S^3$ boundary. The underlying theory living on the $S^3$ boundary is built from the $\mathrm{Sp}(N)$ invariant sector of the $N$ anti-commuting, conformally coupled real scalars $\chi^I$, with $I=1,\ldots ,N$. The glue, indicated by the blue lines, are conformal higher spin gauge fields.}
\label{fig:gluing}
\end{center}
\end{figure}
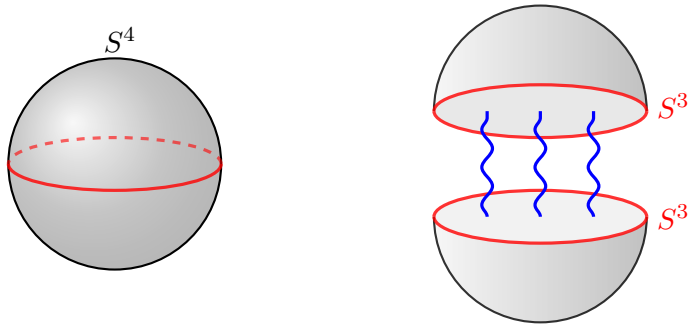

Inspired by the one-loop structure (\ref{oneloophs}), we hypothesise that the full four-sphere partition function of the minimal higher spin model takes the form 
\begin{equation}\label{eq:HS_S4}
\mathcal{Z}_{\mathrm{h.s.}}^{(N)}[S^4] \equiv 
\frac{(-\ii)^{\mathcal{P}}}{\mathrm{vol}\,\mathcal{G}_{\mathrm{HS}}} \int [\mathcal{D}\mathcal{B}]
\Big|\mathcal{Z}_{\mathrm{free}}^{(-N)}[\mathcal{B}]\Big|^2~.
\end{equation}
Pictorially, equation (\ref{eq:HS_S4}) is shown in figure \ref{fig:gluing}. 
\noindent
Here, $\mathcal{Z}_{\mathrm{h.s.}}^{(N)}[S^4]$ denotes the partition function of the full interacting minimal higher spin theory, whereas $\mathcal{Z}_{\mathrm{free}}^{(-N)}[\mathcal{B}]$ is the partition function of the $\text{Sp}(N)$ model, built from  $N$ free anti-commuting and conformally coupled real-valued scalar fields. The reason for switching the statistics of the fields relates to the aforementioned sign difference between the on-shell actions of Euclidean AdS$_4$ and $S^4$ discussed below (\ref{eq:S4}).\footnote{Coupling anti-commuting conformal scalar fields to background sources is loosely reminiscent of the situation in $\Lambda>0$ two-dimensional quantum gravity coupled to a minimal model with large and negative central charge \cite{Anninos:2020geh}.} 
The bilocal function $\mathcal{B}$ encodes a set of sources for the space of $\text{Sp}(N)$ invariant single trace operators. These consist of a scalar and conserved currents for each even spin. For the sake of convenience, as in \cite{Anninos:2017eib},  on $\mathbb{R}^3$ we have 
\begin{equation}\label{eq:B}
    \mathcal{B}(\bold{x},\bold{y}) \equiv b(\bold{x}) \delta_{\bold{x}\bold{y}} + b^{ij}(\bold{x})  \left(\partial_i \partial_j - 3 \overset{\leftarrow}{\partial}_i \partial_j - \delta_{ij} \overset{\leftarrow}{\partial}_l \partial^l  \right) \delta_{\bold{x}\bold{y}}  + \ldots~.
\end{equation}
In the above, the differential operators that appear at each (even) derivative order, are those associated to a traceless and conserved conformal $\mathrm{Sp}(N)$ invariant current built from the $\chi_I$. For instance, we have $\mathcal{O} \equiv \chi^I\chi^J\Omega_{IJ}$ and $\mathcal{O}_{ij} \equiv \chi^I\partial_i \partial_j \chi^J \Omega_{IJ} - 3 \partial_i \chi^I \partial_j \chi^J \Omega_{IJ} + \delta_{ij} \partial_l\chi^I \partial^l\chi^J\Omega_{IJ}$
for the spin-zero and the spin-two operators respectively. (The spin-zero current is strictly speaking not a conserved current, but by slight abuse of terminology we refer to it as such.) As the sources $\{b(\bold{x}),b^{ij}(\bold{x}),\ldots\}$ are conformally covariant, transforming as conformal primaries with the shadow weight of a conformal current, we can always conformally map between $S^3$ and $\mathbb{R}^3$. Also, in (\ref{eq:HS_S4}) we have momentarily absorbed the $N$ dependence of the group volume factor into the measure of $\mathcal{B}$. 
\newline\newline
Explicitly, the fully sourced partition function is given by 
\begin{align}\label{eq:Zfree}
    \mathcal{Z}_{\mathrm{free}}^{(-N)}[\mathcal{B}] 
    &= \int [\mathcal{D}\chi_I]\e^{-\int_{\Omega,\Omega'} \chi^I(\Omega)\left(- \delta(\Omega-\Omega')\nabla_c^2 + \mathcal{B}(\Omega,\Omega') \right) \chi^J(\Omega') \, \Omega_{IJ} }~\cr
    &= \int [\mathcal{D}\chi_I]\e^{-\int_{\Omega} \chi^I(-\nabla_c^2)\chi^J \Omega_{IJ} +b \mathcal{O}+ b^{ij}\mathcal{O}_{ij} + \ldots}~,
\end{align}
where $I=1,\ldots, N$. We have conformally mapped $\mathcal{B}$  from $\mathbb{R}^3$ to $S^3$, and $-\nabla_c^2 \equiv - \nabla^2 + \tfrac{R}{8}$ is the conformal Laplacian on a  unit $S^3$ of Ricci scalar $R=6$. Here the $b_{i_1\ldots i_s}$ are viewed as conformal projections of the $b_{\nu_1 \ldots \nu_s}$ onto an $S^3$ hypersurface in $S^4$. The $\chi_I$ are $N$ free real-valued anti-commuting scalars with conformal dimension $\Delta_{{\chi}_I}= \tfrac{1}{2}$. Finally, the skew symmetric matrix is given by 
\begin{equation}
    \Omega_{IJ} = \begin{pmatrix}
    0 & \mathbb{I}_{\frac{N}{2}}\\
    -\mathbb{I}_{\frac{N}{2}} & 0
    \end{pmatrix}~,
\end{equation}
where $N$ is even. 
We are  using the shorthand notation $\int_{\Omega} \equiv \int \d^3x \sqrt{g}$, where $\Omega$ denotes a point on $S^3$. 
\newline\newline
Crucially, the conformal currents  are all quadratic in the $\chi_I$, so that one can compute the fully sourced partition function
\begin{equation}\label{eq:3dConformally coupled}
    \mathcal{Z}_{\mathrm{free}}^{(-N)}[\mathcal{B}] = {\det}^{\frac{N}{2}}\left(\frac{-\nabla_c^2 + \mathcal{B}}{\Lambda_{\text{u.v.}}}\right) = {\det}^{\frac{N}{2}} \left(\frac{-\nabla_c^2}{\Lambda_{\text{u.v.}}}\right) \e^{\frac{N}{2} \left(\tr \log(\boldsymbol{1} - \nabla_c^{-2}\mathcal{B}) + \tr\nabla_c^{-2}\mathcal{B} \right)} ~.
\end{equation}
Here, $\Lambda_{\text{u.v.}}$ is an ultraviolet cutoff scale required to make sense of the functional determinant,  we have regularised the determinant such that all one-point functions are vanishing, and the trace $\text{tr}$ acts on differential operators.
\newline\newline
One can view $S_{\text{HS}} \equiv -2\log |\mathcal{Z}_{\mathrm{free}}^{(-N)}[\mathcal{B}]|$ as the effective action of an induced conformal higher spin theory in three spacetime dimensions \cite{Basile:2018eac,Segal:2002gd,Beccaria:2014jxa}. The theory is non-local, as it stems from integrating out conformal fields in odd spacetime dimensions. The division by the volume of the three-dimensional conformal higher spin gauge group, $\mathcal{G}_{\mathrm{CHS}}$,  is necessitated by the fact that sources of conformal currents transform as gauge fields \cite{Metsaev:2008fs}. We note that even after gauge-fixing these sources to be transverse and traceless, there is a residual conformal higher spin gauge group given by the constant part of $\mathcal{G}_{\mathrm{CHS}}$. $\mathcal{G}_{\mathrm{CHS}}$ is isomorphic, at the level of the Lie algebra, to the four-dimensional higher spin group, $\mathcal{G}_{\mathrm{HS}}$, much like the dS$_4$ isometries are isomorphic to the three-dimensional conformal symmetries. Thus, in (\ref{eq:HS_S4}) we do not distinguish between $\mathcal{G}_{\mathrm{CHS}}$ and $\mathcal{G}_{\mathrm{HS}}$ in the group volume term, since at least for some choice of real form of the complexified Lie algebra, $\mathcal{G}_{\mathrm{CHS}} \cong \mathcal{G}_{\mathrm{HS}}$. The necessity to divide by the constant part of  $\mathcal{G}_{\mathrm{HS}}$ in (\ref{eq:HS_S4}) is analogous to the fact that one must divide the gravitational sphere path integral by the volume of the de Sitter isometry group (or its Euclidean continuation) after gauge fixing the linearised graviton.

\subsection{Large $N$ expansion}

At large $N$,  expression (\ref{eq:HS_S4}) admits a perturbative expansion in $\tfrac{1}{N}$. To leading order, this yields
\begin{equation}\label{eq:3dConformally coupled ii}
  \log \mathcal{Z}_{\mathrm{free}}^{(-N)}  \approx  \frac{N}{2} \log  {\det} \left(\frac{-\nabla_c^2}{\Lambda_{\text{u.v.}}}\right) = -{N} \int_0^\infty \frac{\d \mathfrak{t}}{2\mathfrak{t}}\frac{1+\e^{-\mathfrak{t}}}{1-\e^{-\mathfrak{t}}}\frac{\e^{-\frac{3\mathfrak{t}}{2}}+\e^{-\frac{t}{2}}}{(1-\e^{-\mathfrak{t}})^2}~.
\end{equation}
On the right hand side, we find the three-sphere partition function of $-N$ free conformally coupled real scalars. Each of them transforms in the complementary series irreducible representation of $\mathrm{SO}(1,3)$, the isometry group of dS$_3$, with $\Delta_+ = \tfrac{3}{2}$ (or equivalently $\Delta_-= \tfrac{1}{2}$). The spatial dimensionality of dS$_3$ is also stressed by the  power of the integrand's denominator in  (\ref{eq:3dConformally coupled ii}).  
\newline\newline
To next order at large $N$, one must expand (\ref{eq:HS_S4}) to quadratic order in $\mathcal{B}$ around the trivial configuration $\mathcal{B}=0$. To compute the path integral over the quadratic fluctuations of $\delta\mathcal{B}$ around the trivial saddle $\mathcal{B}=0$, it is useful to observe that up to quadratic order
\begin{equation}\label{chs2}
\mathcal{Z}_{\mathrm{h.s.}}^{(N)}[S^4]  \approx  {\det}^{\frac{N}{2}} \left(\frac{-\nabla_c^2}{\Lambda_{\text{u.v.}}}\right)\prod_{s \in\text{even}} \int \frac{\left[\mathcal{D}\delta b_{\bold{i}_s}\right]}{{\mathrm{vol}_{{N}}\,\mathcal{G}_{\mathrm{HS}}}} \exp \left({-\frac{1}{2}\int_{\Omega,\Omega'} \delta b_{\bold{i}_s}(\Omega) G^{\bold{i}_s \bold{j}_s}(\Omega,\Omega') \delta b_{\bold{j}_s}(\Omega')} \right)~.
\end{equation}
The above structure follows by recalling the detailed form (\ref{eq:B}) mapped to $S^3$, and we have rescaled the $\delta b_{\bold{i}_s}(\Omega)$ by a factor of $\sqrt{N}$. By $\delta b_{\bold{i}_s}(\Omega)$ we denote the specific perturbative source of the spin-$s$ conformal current,  in the transverse-traceless gauge. It is worth recalling here that the sources of conserved currents are subject to a gauge-redundancy that must be  fixed \cite{Metsaev:2008fs} (see also appendix A of \cite{Anninos:2017eib}). For a spin-$s$ source, the gauge redundancy is sufficient to render the source transverse and traceless. Take as an example the spin-one source, $\delta b_i$. This is redundant under $\delta b_i \to \delta b_i + \partial_i \omega$, where $\omega$ is some smooth function, such that we can always render it transverse. By $G^{\bold{i}_s\bold{j}_s}(\Omega,\Omega')$ we denote the two-point function of the corresponding conserved current. As a simple example, for spin-zero we have
\begin{equation}
G(\Omega,\Omega') =  \frac{1}{N} \langle \mathcal{O}(\Omega)  \mathcal{O}(\Omega') \rangle = \frac{1}{2}\left(-\nabla_c^{-2}\right)\left(-\nabla_c^{-2}\right)~,
\end{equation}
with further explicit examples of higher spin found in \cite{Giombi:2013yva}. One can view (\ref{chs2}) as the free approximation of a three-dimensional conformal higher spin gauge theory \cite{Segal:2002gd,Beccaria:2014jxa,Giombi:2013yva}. When parity even, such a theory is necessarily non-local. 
\newline\newline
Thus, we are instructed in (\ref{chs2}) to compute the functional determinants of higher spin two-point functions. Fortunately, this problem was attacked in \cite{Giombi:2013yva}, where such determinants were computed in a zeta-function regularisation scheme. Moreover, section 9 of \cite{Anninos:2020hfj} shows that there exists a simple expression for the one-loop contribution of a conformal higher spin gauge field, and it takes a character form. Concretely, both the bulk and edge characters of the conformal higher spin field are related to those of  AdS$_4$ (with normalisable boundary conditions) and dS$_4$ as follows
\begin{equation}\label{chschi}
\chi^{(s)}_{\text{HS},{S^d}} =  \chi^{(s)}_{\text{dS}_{d+1}}  - 2 \chi^{(s)}_{\text{AdS}_{d+1}}~.
\end{equation}
As such, the one-loop character contribution on $S^3$ of a spin-$s$ conformal gauge field is
\begin{equation}
\log {Z}^{(s)}_{1\text{-loop,~HS}} = \int_0^\infty \frac{\d\mathfrak{t}}{2\mathfrak{t}} \frac{1+\e^{-\mathfrak{t}}}{1-\e^{-\mathfrak{t}}} \chi^{(s)}_{\text{HS},{S^3}}~.
\end{equation}
Implementing (\ref{chschi}), one finds that only the $s=0$ sector yields a non-trivial contribution, whilst all the higher spin contributions vanish up to zero mode factors. One can understand the latter as an indication that the three-dimensional conformal higher spin fields carry no local degrees of freedom. They do, however, contribute a factor of $N^{-\frac{n_{s,\text{CKT}}}{2}}$ to $\mathrm{vol}_{{N}}\,\mathcal{G}_{\mathrm{HS}}$, where $n_{s,\text{CKT}} = \frac{1}{3}s(4s^2-1)$ is the number of spin-$s$ conformal Killing tensors on $S^3$. This contribution was understood in \cite{Giombi:2013yva,Anninos:2020hfj} to indicate that the volume $\mathrm{vol}_N\,\mathcal{G}_{\mathrm{HS}}$, as dictated by the path integral, should be normalised in units of the coupling constant, which goes as $N^{-\frac{1}{2}}$. Indeed, $n_{s,\text{CKT}}$ counts the number of generators of the higher spin algebra at spin-$s$ \cite{ Eastwood:2002su}.
\newline\newline
The total partition function of the conformal higher spin gauge theory from the character contributions at  one-loop reads
\begin{equation}
   \mathcal{Z}_{1\text{-loop,~HS}}  = \prod_{s \in \text{even}}{Z}^{(s)}_{1\text{-loop,~HS}} =  Z_{\text{HS}}^{(s=0)} ~,
\end{equation}
where the $s=0$ contribution reads
\begin{equation}\label{eq:evalated_HS}
\log Z_{\text{HS}}^{(s=0)} = \int_0^\infty \frac{\d \mathfrak{t}}{2\mathfrak{t}}\frac{1+\e^{-\mathfrak{t}}}{1-\e^{-\mathfrak{t}}}\frac{-\e^{-\mathfrak{t}}}{(1-\e^{-\mathfrak{t}})^2}~.
\end{equation}
Combining (\ref{eq:evaluated_ccscalar}) and (\ref{eq:evalated_HS}), at one-loop and up to divergent terms that can be removed with local counterterms, we have
\begin{align}\label{oneloopN}
 & \frac{1}{\mathrm{vol}\,\mathcal{G}_{\mathrm{HS}}} \int [\mathcal{D}\mathcal{B}] 
\Big|\mathcal{Z}_{\mathrm{free}}^{(-N)}[\mathcal{B}]\Big|^2 \approx \frac{1}{\mathrm{vol}_{{N}}\,\mathcal{G}_{\mathrm{HS}}} \times \e^{2{N} \left( \frac{1}{8}\log 2 - \frac{3\zeta(3)}{16\pi^2} \right) +\frac{\zeta(3)}{8\pi^2}} ~.
\end{align}
This is in line with the computation of the higher spin four sphere partition function (\ref{oneloopN_4D}), thus confirming (\ref{eq:HS_S4}) to one-loop order. In principle, one can proceed to calculate higher loop corrections also, which are encoded in the $\tfrac{1}{N}$ expansion. The higher loop terms will require a definition of the measure $[\mathcal{D}\mathcal{B}]$; such a measure was proposed in \cite{Anninos:2017eib}.
\newline\newline
We should emphasise, however, that the expression (\ref{eq:HS_S4}) remains somewhat formal due to three distinct divergences. Firstly, there are ultraviolet divergences stemming from the path integrals over the various fields. Perhaps these can be dealt with by adding local counterterms, though we should remember that such counterterms would be local in three rather than four dimensions.  Secondly, the dimension of the conformal higher spin group itself, $\text{dim} \,\mathcal{G}_{\mathrm{HS}}=\sum_s n_{s,\mathrm{CKT}}$, diverges. Thirdly, the volume of the conformal higher spin group does not have a known mathematical definition. We could also add to this list the indefiniteness of $\mathcal{P}$. In section \ref{sec5}, we will discuss a supersymmetric model that improves these issues. Before doing so, we will discuss one more potential interpretation of the gluing formula (\ref{eq:HS_S4}).

\section{dS$_4$ wavefunction perspective} \label{sec4}

In this section, we would like to offer a complementary perspective on
(\ref{eq:HS_S4}). The quadratic structure indicates a bilinear pairing between
two conformally invariant partition functions. From this viewpoint, one may be
tempted to propose a wavefunction interpretation
\begin{equation}\label{eq:Z HH}
   \mathcal{Z}_{\mathrm{free}}^{(-N)}[\mathcal{B}] = \Psi_{\mathrm{HH}}[\mathcal{B}]~,
\end{equation}
with $\mathcal{Z}_{\mathrm{free}}^{(-N)}[\mathcal{B}]$ given by  (\ref{eq:3dConformally coupled}), and
$\Psi_{\mathrm{HH}}$ is the Hartle-Hawking wavefunction of the higher spin theory on an $S^3$ spatial slice.
This is reminiscent of the higher spin dS$_4$/CFT$_3$ correspondence proposed
and explored in
\cite{Anninos:2011ui,Anninos:2012ft}, following the general
ideas of
\cite{Strominger:2001pn,Witten:2001kn,Maldacena:2002vr}.
\newline\newline
In the usual dS/CFT picture, one considers the Hartle-Hawking wavefunction (or some other physical quantity) as a functional of the conformal boundary data at the infinite future, $\mathcal{I}^+$. In the higher spin dS$_4$/CFT$_3$ correspondence, the bulk fields are associated to composite operators built from the free boundary degrees of freedom, $\chi_I$, of the $\text{Sp}(N)$ model. For instance, given the free field $\chi^I$ of conformal dimension $\Delta_{\chi^I}=\frac{1}{2}$, one can build the composite operator $\mathcal{O} = \chi^I \chi^J \Omega_{IJ}$, with $\Delta_{\mathcal{O}}=1$, that is dual to the boundary profile of the bulk scalar at $\mathcal{I}^+$. A similar relation holds for the higher spin currents $\mathcal{O}_\bullet$, which are all quadratic in the $\chi^I$. Thus, although the boundary theory is free, the mapping into composite operators induces non-trivial interactions from the bulk point of view. That conformal structures live at $\mathcal{I}^+$ reflects the property of de Sitter isometries projected onto the future boundary where they act as conformal transformations of the boundary metric.
\newline\newline
We can contrast the above more customary dS/CFT picture with the gluing formula (\ref{eq:HS_S4}).  In the latter, we are no longer anchoring the conformal fields $\mathcal{B}$ at $\mathcal{I}^+$. In fact, the gluing formula was motivated entirely from considerations of the bulk theory on a finite volume $S^4$! A priori, there is no reason why a conformal structure such as $\mathcal{B}$  should reside on a finite size hypersurface  within the four-sphere. Our interpretation is that (\ref{eq:HS_S4}) implements conformal boundary conditions on the finite-size spatial hypersurface, rather than at $\mathcal{I}^+$. Such boundary conditions fix the conformal three-metric, $\gamma_{ij}$, and the trace of the extrinsic curvature, $\mathrm{tr}K(\Omega)$, along the finite size hypersurface. This is reminiscent of the conformal boundary condition in Euclidean general relativity, which has been argued to be mathematically well-defined \cite{Anderson:2006lqb}, and studied in the context of $\Lambda>0$ theories in \cite{Anninos:2024wpy,Banihashemi:2025qqi,Anninos:2025fer}. From this point of view, enforcing that the spatial hypersurface becomes the infinite volume $\mathcal{I}^+$ corresponds to tuning $\mathrm{tr}K$ to a special complex value. 
\newline\newline
From the wavefunction perspective, (\ref{eq:Z HH}), the full four-sphere path integral is viewed as the norm of the Hartle-Hawking wavefunction of the higher spin theory
\begin{equation}\label{eq: Norm HH}
 \langle \Psi_{\mathrm{HH}} | \Psi_{\mathrm{HH}} \rangle \overset{?}{=}\frac{1}{\mathrm{vol}\mathcal{G}_{\mathrm{CHS}}} \int [\mathcal{D}\mathcal{B}] |\mathcal{Z}^{(-N)}_{\mathrm{free}}[\mathcal{B}]|^2
 = \mathcal{Z}_{\mathrm{h.s.}}^{(N)}[S^4]~.
\end{equation}
Interpreted as a norm, (\ref{eq:HS_S4})  suggests that the argument of the wavefunctional $\Psi_{\mathrm{HH}}$ is only path integrated over the conformal structure --- now extended to the space of higher spin conformal sources $\mathcal{B}$ --- while the higher spin extension of $\mathrm{tr} K$ is kept fixed. From this perspective, the fact that the inner product is independent of the York time \cite{York:1972sj}, $\mathrm{tr} K$, can be viewed as the condition of a time-preserving probability. Circumstantial evidence for such a pairing was observed in the context of two-dimensional quantum gravity \cite{Anninos:2025fer}, where $\mathrm{tr}K$ was
shown to drop out of a bilinear pairing of two disk partition functions, each non-trivially depending on $\mathrm{tr} K$, subject to conformal boundary conditions. A related discussion, formulated in terms of an inner product at $\mathcal{I}^+$ and for ordinary gravity, appears in \cite{Collier:2025lux, Cotler:2025gui}. In \cite{Cotler:2025gui} one would replace the residual conformal higher spin gauge group, $\mathcal{G}_{\mathrm{CHS}}$, with some real form of the complexification of $\mathrm{SO}(1,4)$. The discussion is also reminiscent of a Euclidean version of the dS/dS correspondence \cite{Alishahiha:2004md}.
\newline\newline
Generally speaking, it is unclear whether the gravitational sphere path integral can be interpreted as a norm due to the presence of an overall phase that renders it complex \cite{Polchinski:1988ua}. This phase arises from a complex deformation of the path-integration contour for the conformal mode \cite{Gibbons:1978ac}, and differs between different saddles \cite{Anninos:2025ltd}. In the higher spin theory, one has a one-loop phase $\mathcal{P}$ (\ref{totP}) that needs to be regularised somehow.\footnote{In the next section, we will find a less ambiguous phase for an $\mathcal{N}=2$ supersymmetric model. Also, for the non-minimal bosonic higher spin theory, it has been argued that the phase $\mathcal{P}$  vanishes upon regularization \cite{Giombi:2026sqa}.} Had one insisted on a real integration contour for the conformal mode, the gravitational path integral would diverge due to the unboundedness of the Euclidean action. This behaviour should be contrasted with that of a quantum field theory defined on a fixed de Sitter background, where the sphere path integral is indeed the norm of the Euclidean vacuum state. That being said, the pairing structure of  (\ref{eq:HS_S4}) over the conformal sources remains evocative. 
\newline\newline
As a final remark, in de Sitter space all symmetries are a priori gauged \cite{Higuchi:1991tk}. The path integral (\ref{eq:HS_S4}) is one of a conformal higher spin gauge theory, and consequently we must divide by the volume of any residual gauge group, $\mathrm{vol}\,\mathcal{G}_{\mathrm{HS}}$. As previously mentioned, the Euclidean gravity path integral is rendered finite by complexifying the path-integration contour \cite{Gibbons:1978ac}. Doing so raises the question --- how do we select a real form for the complexified residual diffeomorphism group whose volume we should divide by? For $\Lambda>0$, natural candidates are $\mathrm{SO}(5)$ and $\mathrm{SO}(1,4)$. (Dividing by a non-compact group such as $\mathrm{SO}(1,4)$ which has infinite volume is subtle, particularly in the Euclidean setting.) 
Along a similar vein, one might also consider complexifying the cutoff parameter required to regularise the partition functions in (\ref{eq:HS_S4}). If so, local counterterms will generally have complex coefficients. The resulting partition function $\mathcal{Z}^{(-N)}_{\mathrm{free}}[\mathcal{B}]$ would then exhibit oscillatory properties, a feature that is  suggestive  of the oscillatory behaviour of a de Sitter wavefunction.

\section{$\mathcal{N}=2$ super-gluing formula}\label{sec5}

In this section, we generalise the previous story to the supersymmetric non-minimal higher spin model discussed in \cite{Sezgin:2012ag,Hertog:2017ymy}. The theory we will consider has bulk field content given by a conformally coupled complex scalar, a spin-$\tfrac{1}{2}$ massless Dirac fermion, and a complex totally massless gauge field at every spin $s\in\{1,\tfrac{3}{2},2,\frac{5}{2},\ldots\}$. The model thus realises the whole collection of totally massless discrete series unitary irreducible representations of $\mathrm{Spin}(1,4)$. There is a Euclidean AdS$_4$ counterpart to this theory \cite{Leigh:2003gk,Sezgin:2012ag,Chang:2012kt,Lang:2024dkt}, which is dual to a free superconformal $U(N)$ vector model with $\mathcal{N}=2$ supersymmetry. By the standard AdS/CFT dictionary, for each bulk field there is a corresponding $U(N)$ invariant (super-)conformal current. We now study expression (\ref{eq:HS_S4}) generalised to the supersymmetric case. The possibility of a supersymmetric de Sitter theory may appear as a surprise \cite{Pilch:1984aw,Lukierski:1984it}. We should note, however, as emphasised in recent work \cite{Letsios:2023qzq,Anninos:2023exn,Anninos:2025mje,Chen:2025foq, Higuchi:2025pbc,Boulanger:2026wnw}, that higher spin fermionic fields furnish unitary irreducible representations of the de Sitter group $\mathrm{SO}(1,4)$, and may provide the building blocks for a reasonable interacting theory, such as the higher spin theories explored in \cite{Sezgin:2012ag,Hertog:2017ymy}. In any case, we can certainly consider the supersymmetric extension of (\ref{eq:HS_S4}) and see where it leads us.

\subsection{A prescription for the $S^4$ on-shell action}

As in the bosonic case, we lack access to the on-shell action of the supersymmetric dS$_4$ higher spin theory. We will propose an on-shell action for the dS$_4$ higher spin theory from the AdS$_4$ counterpart, which is dual to the free $\mathcal{N}=2$ superconformal $U(N)$ vector model.\footnote{As before, a more complete treatment would require coupling to an $\mathcal{N}=2$ supersymmetric Chern-Simons theory with $U(N)$ gauge group at parameterically large level. On an $S^3$, the two sectors decouple, however the $S^3$ partition function is multiplied by a Chern-Simons piece also.} Since the $\mathcal{N}=2$ theory is simply a collection of $N$ free Dirac fermions and $N$ free complex scalars, now with ordinary statistics, to compute its three-sphere partition function we can implement the results for the determinants, presented later in section \ref{largeNsg}. One finds 
\begin{equation}\label{logAdS}
\log \mathcal{Z}_{\mathcal{N}=2}[S^3] = - \frac{N}{2}\log 2~.
\end{equation}
In the above, we have removed divergences going as $\mathcal{O}(\frac{1}{\varepsilon})$ with local counterterms, while the $\mathcal{O}(\frac{1}{\varepsilon^3})$ terms naturally cancel between the bosons and fermions.
\newline\newline
The dual AdS$_4$ bulk field content is given by two conformally coupled real scalars, a spin-$\tfrac{1}{2}$ massless Dirac fermion, and a complex totally massless gauge field at every spin $s\in\{1,\tfrac{3}{2},2,\frac{5}{2},\ldots\}$. Such bulk theories were studied in \cite{Chang:2012kt,Sezgin:2012ag,Leigh:2003gk}. One of the scalars is quantised with the standard boundary conditions, and the other with the alternate boundary conditions reflecting the fact that the two $U(N)$ invariant scalars in the dual CFT have conformal weights $\Delta=1$ and $\Delta=2$. Given the bulk spectrum, we can compute the one-loop contribution to the EAdS$_4$ partition function. The sum over the fermionic tower, remarkably, vanishes on its own (see also (6.32) of \cite{Anninos:2025mje}):
\begin{equation}
\sum_{s\in\{\frac{1}{2},\frac{3}{2},\ldots\}} \left(\chi^{(s)}_{\mathrm{bulk}}(\mathfrak{t})-\chi^{(s)}_{\mathrm{edge}}(\mathfrak{t}) \right)  = 0~.
\end{equation}
We are then left with a sum over the bosonic tower, yielding the character sum
\begin{equation}\label{SAdS1loop}
\log \mathcal{Z}^{\text{one-loop}}_{\text{AdS}_4} =  \int_0^\infty \frac{\d\mathfrak{t}}{2\mathfrak{t}} \frac{1+\e^{-\mathfrak{t}}}{1-\e^{-\mathfrak{t}}} \left( \frac{\e^{-\mathfrak{t}}+\e^{-2\mathfrak{t}}}{(1-\e^{-\mathfrak{t}})^3} + 2 \sum_{s=1}^\infty\left(\chi^{(s)}_{\mathrm{bulk}}(\mathfrak{t})-\chi^{(s)}_{\mathrm{edge}}(\mathfrak{t})\right)\right)~.
\end{equation}
We would like to emphasize that (compared to the de Sitter characters), the first character is really the sum of two anti-de Sitter characters, one for a scalar with conformal weight $\Delta=1$ and one with conformal dimension $\Delta=2$. Curiously, the total EAdS$_4$ one-loop partition function (\ref{SAdS1loop}) is precisely one-half of the analogous calculation for the $\mathcal{N}=2$ higher spin theory on $S^4$. Concretely, performing the sum we are left with 
\begin{equation}\label{SAdS1loop2}
\log \mathcal{Z}^{\text{one-loop}}_{\text{AdS}_4} = -\int_0^\infty \frac{\d\mathfrak{t}}{2\mathfrak{t}} \frac{1+\e^{-\mathfrak{t}}}{1-\e^{-\mathfrak{t}}} \frac{\e^{-\mathfrak{t}}}{(1-\e^{-\mathfrak{t}})^2}  = +\frac{\zeta(3)}{8\pi^2}~,
\end{equation}
where we have again removed divergences with boundary local counterterms. Similarly, for the same spectrum in four-dimensional Euclidean de Sitter space we have the sum
\begin{equation}
\log \mathcal{Z}^{\text{one-loop}}_{\text{dS}_4} = -2\int_0^\infty \frac{\d\mathfrak{t}}{2\mathfrak{t}} \frac{1+\e^{-\mathfrak{t}}}{1-\e^{-\mathfrak{t}}} \frac{\e^{-\mathfrak{t}}}{(1-\e^{-\mathfrak{t}})^2}  = +\frac{\zeta(3)}{4\pi^2}~.
\end{equation}
In order for the on-shell Euclidean AdS$_4$ action plus the one-loop contribution to agree with the dual CFT prediciton (\ref{logAdS}), we postulate that (up to divergences that can be eliminated leaving an unambiguous finite part) the on-shell action is given by
\begin{equation}\label{SUSYAdS}
-S^{(N)}_{\text{AdS}_4} = -\frac{N}{2}\log 2 - \frac{\zeta(3)}{8\pi^2}~.
\end{equation}
Based on this, and following the discussion near (\ref{eq:S4}), we hypothesise that the corresponding four-sphere on-shell action is 
\begin{equation}\label{SUSYdS}
-S^{(N)}_{\text{dS}_4} = -2  S^{(-N)}_{\text{AdS}_4} =  {N}\log 2 - \frac{\zeta(3)}{4\pi^2}~.
\end{equation}
It is curious that the supersymmetric theory has a non-vanishing one-loop contribution (\ref{SAdS1loop2}), and that the action (\ref{SUSYAdS}) is corrected in a way that does not correspond to a simple shift in $N$. Perhaps it is a factor that we must account for so as to have a well-defined measure. One method to sharpen this would be to compute the relation of $N$ to $G_N\Lambda$ from other observables, such as AdS$_4$ boundary three-point functions \cite{Giombi:2009wh}.

\subsection{A super-gluing formula}

Inspired by (\ref{eq:HS_S4}), we  consider its natural $\mathcal{N}=2$ generalisation
\begin{equation}\label{eq:HS_S4_N2}
{
    \mathcal{Z}_{\mathcal{N}=2\,\mathrm{h.s.}}^{(N)}[S^4] \equiv \frac{1}{\mathrm{vol} \,\mathcal{G}_{\mathrm{sHS}}} \int [\mathcal{D}\mathcal{B}_{s}] \Big|\mathcal{Z}_{\mathrm{free}}^{(-N)}[\mathcal{B}_{s}]\Big|^2}~,
\end{equation}
where now $\mathcal{B}_{s}$ denotes the space of sources for all $U(N)$ invariant (super-)currents. Here, $\mathcal{Z}_{\mathrm{free}}^{(-N)}[\mathcal{B}_s]$ is the partition function of $N$ free anti-commuting complex scalars $\chi^I$ and $N$ free spin-$\frac{1}{2}$ commuting Dirac fermions with a general source turned on for all $U(N)$ invariant (super-)currents. 
\newline\newline
As already mentioned, the partition function, $\mathcal{Z}_{\mathrm{free}}^{(-N)}[\mathcal{B}_{s}]$, is invariant under the $\mathcal{N}=2$ supersymmetric higher spin symmetries.  
The free $\mathcal{N}=2$ supersymmetric $U(N)$ invariant action on a round $S^3$ is given by
\begin{equation}\label{freeN2S}
S_{\mathcal{N}=2} =  \int_{\Omega} \left( g^{ij} \partial_i \tilde{\chi}_I \partial_j \chi_I + \frac{R}{8} \tilde{\chi}_I \chi_I - \ii \tilde{\psi}_I\gamma^i D_i \psi_I + \widetilde{F}_I F_I \right)~,
\end{equation}
where now the $\chi_I$ denote $N$ complex valued anti-commuting scalars, and $\psi_{I}$ denote $N$ spin-$\tfrac{1}{2}$ Dirac commuting fermions.  We have also included auxiliary anti-commuting scalars,  ${F}_I$, which can always be set to zero on-shell. The three-dimensional Euclidean $\gamma$-matrices satisfy the Clifford algebra $\{\gamma_i,\gamma_j\} = 2g_{ij}$.  Again, the switching of the statistics parallels the previous section and is related to the $N\rightarrow -N$ sign difference between the on-shell Euclidean AdS$_4$ and Euclidean dS$_4$ actions. 
\newline\newline
In addition to the $U(N)$ symmetry, the above model enjoys an $\mathcal{N}=2$ superconformal symmetry. On $\mathbb{R}^3$, the global $\mathcal{N}=2$ supersymmetric variations (see for example the reviews \cite{Hama:2010av,Willett:2016adv}) are 
\begin{equation}
\label{eq:SuperConformalTransformations}
\delta \chi_I = -\tilde{\epsilon}\psi_I ~, \quad\quad 
\delta \psi_I =  \ii \gamma^i \epsilon \partial_i \chi_I  + \tilde{\epsilon}F_I  ~,  \quad\quad
\delta F_I = \ii {\epsilon} \gamma^i \partial_i \psi_I ~,
\end{equation}
and similarly for the tilded fields. The supersymmetry transformations on $S^3$ can be obtained by a conformal map.
The above transformations account for the reversed statistics of the fields:  $\chi_I$ and $F_I$ are anti-commuting fields, while $\psi_I$ is commuting, and similarly for the tilded fields. 
 Here $\epsilon$ and $\tilde{\epsilon}$ are conformal Killing spinors on $\mathbb{R}^3$, or $S^3$ upon conformally mapping, and satisfy $D_i \epsilon = \tfrac{\gamma_i}{3} \slashed{D}\epsilon$. We have suppressed spinorial indices in the above. The bosonic subalgebra of the three-dimensional Euclidean $\mathcal{N}=2$ superconformal algebra is $\mathfrak{so}(1,4)\times \mathfrak{u}(1)_R$, consisting of eleven bosonic generators, which are accompanied by eight fermionic generators. 
\newline\newline
It will prove useful to describe the theory in terms of the $\mathcal{N}=2$ superspace formalism, whereby in addition to the three spacetime coordinates, $x^i$, the fields  depend also on Grassmann coordinates $(\theta_\alpha,\tilde{\theta}_\alpha)$ with $\alpha=\{1,2\}$, and $\theta_\alpha\theta_\beta=-\theta_\beta\theta_\alpha$ and so on. We have that $\theta^2 \equiv \theta_\alpha \theta^\alpha$, where $\epsilon^{12}=+1$ following the general rule that spinorial indices are raised with an anti-symmetric tensor $\epsilon^{\alpha\beta}$. It is convenient to describe the theory on $\mathbb{R}^3$, whereby the supercharges and the covariant derivatives in superspace 
take the simple form
\begin{equation}
Q_\alpha \equiv \frac{\partial}{\partial{\theta^\alpha}} - \ii (\gamma^i  \tilde{\theta})_\alpha \partial_i \quad\quad \text{and} \quad\quad \widetilde{Q}_\beta \equiv -\frac{\partial}{\partial{\tilde{\theta}^\beta}} + \ii ( \gamma^i\theta  )_\beta \, \partial_i~, 
\end{equation}
\begin{equation}
 D_\alpha \equiv \frac{\partial}{\partial{\theta^\alpha}} + \ii (\gamma^i  \tilde{\theta})_\alpha \partial_i \quad\quad \text{and} \quad\quad \widetilde{D}_\beta \equiv -\frac{\partial}{\partial{\tilde{\theta}^\beta}} - \ii  ( \gamma^i{\theta}  )_\beta \, \partial_i~.
\end{equation}
We then have the non-trivial anti-commutation relation $\{Q_\alpha,\widetilde{Q}_\beta\} = 2\ii (\gamma^i)_{\alpha\beta}\partial_i$. 
A chiral superfield $\widetilde{D}_\beta \Phi=0$ is given by 
\begin{equation}
 \Phi(z) = \chi(y^i) + \sqrt{2}\theta \psi(y^i) +\theta^2 F(y^i)~,   
\end{equation}
where the chiral coordinate $y^i \equiv x^i + \ii \theta \gamma^i \tilde{\theta}$ is annihilated by $\widetilde{D}_\alpha$, and $z\equiv (y^i,\theta)$.
Similarly, an anti-chiral superfield (${D}_\alpha \widetilde{\Phi}=0$) is given by $\widetilde{\Phi}(\tilde{z}) = \tilde{\chi}(\tilde{y}^i) - \sqrt{2}  \tilde{\psi}(\tilde{y}^i) \tilde{\theta} +  \tilde{\theta}^2 \widetilde{F}(\tilde{y}^i)$, with $\tilde{y}^i \equiv x^i - \ii \theta \gamma^i \tilde{\theta}$ being annihilated by ${D}_\alpha$. The free $\mathcal{N}=2$ action of a chiral multiplet on $\mathbb{R}^3$, expressed in the superspace language, reads 
\begin{equation}\label{freeN2theta}
 S_{\mathcal{N}=2} = \int \d z \,\widetilde{\Phi}_I(\tilde{z})\Phi_I(z)~,\quad \d z\equiv \d^3x\d^2 \theta \d^2 \tilde{\theta}~.
\end{equation}
The theory on $S^3$ is obtained by a conformal transformation. We note that (\ref{freeN2theta}) is an integral of a real superfield over all of the superspace coordinates, and hence what is often referred to as a $D$-term. This  indicates that it is $Q$-exact, since integration over the Grassmann valued $\theta$ and $\tilde{\theta}$ amounts to differentiating  the real superfield $\widetilde{\Phi}(\tilde{y}^i,\tilde{\theta})\Phi(y^i,\theta)$ with respect to $\theta$ and $\tilde{\theta}$ (assuming that we can drop boundary terms at spatial infinity) and this in turn amounts to acting with the superspace representation of all supercharges.
The supersymmetry variations (\ref{eq:SuperConformalTransformations}) can be extracted by acting on the chiral and anti-chiral superfields with $Q_\alpha$, $\widetilde{Q}_\beta$.

\subsection{The space of supersymmetric sources}

The space of $U(N)$ invariant sources, which we denote by $\mathcal{B}_{s}$, is now extended by contracting the $\chi_I$ fields with the $\tilde{\psi}_I$ fields and so on. The corresponding partition function is schematically given by
\begin{equation}
    \mathcal{Z}_{\mathrm{free}}^{(-N)}[\mathcal{B}_s] = \int [\mathcal{D}\tilde{\chi}_I][\mathcal{D}\chi_I][\mathcal{D}\tilde{\psi}_{ I}][\mathcal{D}\psi_{ I}]\e^{-S_{\mathcal{N}=2} +\int_{\Omega} \left( b_{\tilde{\chi}\chi}\tilde{\chi}_I \chi_I + b_{\tilde{\psi}\psi}\tilde{\psi}_I \psi_I  + b_{\tilde{\chi}{\psi}} \tilde{\chi}_I \psi_I+b_{\tilde{\psi}{\chi}}\tilde{\psi}_I \chi_I + \ldots \right)}~.
\end{equation}
The set of sources reflects the bulk spectrum of the $\mathcal{N}=2$ higher spin dS$_4$ theory --- a complex bulk field at each spin $s\in\{0,\tfrac{1}{2},1,\tfrac{3}{2},2,\ldots\}$. It is convenient to describe the complete space of sources for $U(N)$ invariant operators in the superspace formalism. Again, for simplicity, we will work on $\mathbb{R}^3$, recalling that we can always use conformal covariance to map back to $S^3$. Consider a general bilocal superfield $\mathcal{B}_s(z_1;\tilde{z}_2)$;  the contribution to the action from the source term is given by
\begin{equation}\label{supersource}
S_{b_s} = \int\d z_1 \d z_2 \, \widetilde{\Phi}_I(\tilde{z}_1) \mathcal{B}_s(z_1;\tilde{z}_2) {\Phi}_I(z_2)~.
\end{equation}
The above action is itself invariant under $\mathcal{N}=2$ supersymmetry provided that the bilocal superfield $\mathcal{B}_s$ satisfies $\widetilde{D}^{(1)}_{\beta}\mathcal{B}_s = 0 =  D^{(2)}_{\alpha}\mathcal{B}_s $, i.e. is chiral with respect to $z_1 \equiv (y_1,\theta)$ and anti-chiral with respect to $\tilde{z}_2 \equiv (\tilde{y}_2,\tilde{\eta})$. In particular, this implies that the superfield $\mathcal{B}_s$ is independent of $\tilde{\theta}$ and $\eta$, and can be written as
\begin{multline}
    \mathcal{B}_s(z_1;\tilde{z}_2)= B_{\tilde{\chi}\chi} (y_1 ; \tilde{y}_2) + \theta B_{\tilde{\psi}\chi}(y_1;\tilde{y}_2) + \tilde{\eta} {{B}}_{\tilde{\chi}\psi}(y_1;\tilde{y}_2) + \theta^2 B_{\chi\widetilde{F}}(y_1;\tilde{y}_2) + \tilde{\eta}^2 {B}_{\tilde{\chi}F}(y_1;\tilde{y}_2)  \cr
    + \theta\tilde{\eta} B_{\tilde{\psi}\psi}(y_1;\tilde{y}_2) +\theta^2 \tilde{\eta} B_{\psi\widetilde{F}}(y_1;\tilde{y}_2) +  \tilde{\eta}^2\theta B_{\tilde{\psi}F}(y_1;\tilde{y}_2) +  \theta^2 \tilde{\eta}^2 B_{\widetilde{F}F}(y_1;\tilde{y}_2)~.
\end{multline}
Bilocal superfields also appear in supersymmetric SYK models (see e.g. \cite{Yoon:2017gut}). At this stage, our task is to open up (\ref{supersource}) and ensure that all $U(N)$ invariant bilocal operators are sourced by $\mathcal{B}_s(z_1;\tilde{z}_2)$. This amounts to extracting the top form in the Grassmann coordinates, as this is the only one that survives the Grassmann integral. For instance, $b_{\widetilde{F}F}(x_1;x_2)\equiv B_{\widetilde{F}F}(x_1;x_2)$ is readily found to be the source of $\widetilde{F}_I(x_1)F_I(x_2)$, whilst $b_{\tilde{\chi}\chi}(x_1;x_2) \equiv\nabla^2_1 \nabla^2_2 B_{\tilde{\chi}\chi} (x_1 ; x_2)$ sources $\tilde{\chi}_I(x_1)\chi_I(x_2)$. More generally, we find the following sources
\[
\renewcommand{\arraystretch}{1.25}
\begin{array}{@{}l@{\quad}l@{\qquad}l@{\quad}l@{}}
b_{\tilde{\chi}\chi}(x_1;x_2)
& \equiv \nabla_1^2\nabla_2^2 B_{\tilde{\chi}\chi}(x_1;x_2)~,
&
b_{\widetilde{F}\psi}(x_1;x_2)
& \equiv 
  \slashed{\partial}_{2} B_{\widetilde{F}\psi}(x_1;x_2)~, \\

b_{\widetilde{F}F}(x_1;x_2)
& \equiv B_{\widetilde{F}F}(x_1;x_2)~,
&
b_{\tilde{\psi}F}(x_1;x_2)
& \equiv 
  \slashed{\partial}_{1} B_{\tilde{\psi}F}(x_1;x_2)~, \\

b_{\tilde{\chi}F}(x_1;x_2)
& \equiv \nabla_1^2 B_{\tilde{\chi}F}(x_1;x_2)~,
&
b_{\widetilde{F}\chi}(x_1;x_2)
& \equiv \nabla_2^2 B_{\widetilde{F}\chi}(x_1;x_2)~, \\

b_{\tilde{\psi}\psi}(x_1;x_2)
& \equiv 
  \slashed{\partial}_{1}\slashed{\partial}_{2}
  B_{\tilde{\psi}\psi}(x_1;x_2)~,
&
b_{\tilde{\chi}\psi}(x_1;x_2)
& \equiv 
  \slashed{\partial}_{2}\nabla_1^2
  B_{\tilde{\chi}\psi}(x_1;x_2)~, \\

b_{\widetilde{\psi}\chi}(x_1;x_2)
& \equiv 
  \slashed{\partial}_{1}\nabla_2^2
  B_{\tilde{\psi}\chi}(x_1;x_2)~.
& &
\end{array}
\]
In the above we have chosen a specific normalisation for the sources, which can be viewed as a normalisation choice for the $U(N)$ invariant operators. The subscript in the derivatives labels the coordinates they act on. A convenient way to test the above expressions is to take $\mathcal{B}_s(z_1;\tilde{z}_2) = {\mathcal{Q}}(z_1) \widetilde{\mathcal{Q}}(\tilde{z}_2)$  to be the product of a chiral and anti-chiral superfield, such that (\ref{supersource}) becomes a product of two free actions (\ref{freeN2theta}). Our bilocal superfield $\mathcal{B}_s$ indeed sources all of the $U(N)$ invariant bilocal operators.
\newline\newline
\noindent \textbf{$\mathcal{N}=2$ higher spin superalgebra.} The higher spin supersymmetric transformations are those superspace linear maps of the chiral and anti-chiral fields that leave the free action invariant. They have been studied in \cite{vasiliev1988extended}. These, in turn, induce a transformation on the superfield sources $\mathcal{B}_s$, which when expanded out  into its component fields will yield an elaborate family of higher spin supersymmetric transformations generalising those of the purely bosonic theory. The partition function of the theory as a function of $\mathcal{B}_s$ will be invariant under these higher spin supersymmetric transformations, effectively defining a three-dimensional non-local superconformal higher spin theory. As for the bosonic theory, the higher spin superalgebra will have an infinite number of bosonic and fermionic generators, and the number of these at spin-$s$ is $\frac{2}{3}s(4s^2-1)$ with $s\in\{1,\tfrac{3}{2},2,\ldots\}$. The $\mathcal{N}=2$ superconformal algebra is a finite dimensional subalgbebra of the $\mathcal{N}=2$ higher spin superalgebra, which forms the basic seed of the full higher spin algebra.

\subsection{Large $N$ expansion of $\mathcal{N}=2$ super-gluing formula}\label{largeNsg}
Again, we can study the partition function $\mathcal{Z}_{\mathrm{free}}^{(-N)}[\mathcal{B}_{s}]$ in a large-$N$ expansion. To leading order, we can set the sources $\mathcal{B}_{s} = 0$ and obtain the expression
\begin{equation}\label{Zsusy}
    \mathcal{Z}_{\mathrm{free}}^{(-N)}[\mathcal{B}_{s}] \approx \frac{{\det}^{{N}}(-\nabla_c^2)}{{\det}^{N} \,\slashed{D}}~, 
\end{equation}
where we have defined the Dirac operator $\slashed{D} \equiv i \gamma^i D_i$, and have suppressed the dependence on $\Lambda_{\text{u.v.}}$. Evaluating the functional determinants in a heat-kernel regularisation scheme (see, for example, appendix C of \cite{Anninos:2020hfj}) yields 
\begin{eqnarray}
-\log {\det} \, \slashed{D} &=& \frac{1}{4}\log 2 + \frac{3\zeta(3)}{8\pi^2} +\frac{\pi}{\varepsilon^3} -\frac{\pi}{4\varepsilon}~, \label{logdetb} \\
\log {\det}(-\nabla_c^2) &=&  \frac{1}{4}\log 2 -  \frac{3\zeta(3)}{8\pi^2} - \frac{\pi}{\varepsilon^3} -\frac{\pi}{8\varepsilon}~. \label{logdetf}
\end{eqnarray}
The $\mathcal{O}\left(\tfrac{1}{\varepsilon^3}\right)$ divergences cancel between the two determinants due to the supersymmetry of the problem. 
\newline\newline
It is worth pausing here for a curious observation. We can express the ratio of determinants (\ref{Zsusy}) in the Harish-Chandra character language of \cite{Anninos:2020hfj}, whereby
\begin{equation}\label{wg}
\log \frac{{\det}(-\nabla_c^2)}{{\det} \,\slashed{D}} = -2 \int_{\mathbb{R}^+} \frac{\d \mathfrak{t}}{2\mathfrak{t}} \left(\frac{1+q}{1-q}\frac{q^{\frac{1}{2}}+q^{\frac{3}{2}}}{(1-q)^2}-\frac{2 q^{\frac{1}{2}}}{1-q} \frac{2 q}{(1-q)^2} \right) =  \int_{\mathbb{R}^+} \frac{\d \mathfrak{t}}{2\mathfrak{t}} \frac{-2q^{\frac{1}{2}}}{1-q}~,\quad q\equiv \e^{-\mathfrak{t}}~.
\end{equation}
The last equality resembles the partition function of a quantum mechanics coupled to worldline gravity. Alternatively, using equation (2.8) of \cite{Anninos:2020hfj}, we note that the square of (\ref{Zsusy}) yields
\begin{equation}\label{worldline}
\exp 2N\int_{\mathbb{R}^+} \frac{\d \mathfrak{t}}{2\mathfrak{t}} \frac{-2q^{\frac{1}{2}}}{1-q} = \lim_{\beta^+\to 0} \left( \e^{\frac{\beta \omega}{2}} +  \e^{-\frac{\beta\omega}{2}}\right)^N =  \lim_{\beta^+\to 0}  \left(Z_{\text{fermion}}[\beta]\right)^N = 2^N~.
\end{equation}
The small-$\mathfrak{t}$ divergence on the left hand side is regularised by splitting the double pole at $\mathfrak{t}=0$ as $\mathfrak{t}^{-2} \to \tfrac{1}{2}\left((\mathfrak{t}+i\varepsilon)^{-2}+(\mathfrak{t}-i\varepsilon)^{-2} \right)$. This is the partition function of $N$ free quantum mechanical fermions at infinite temperature --- a feature endemic of a maximally mixed state.
\newline\newline
Combining (\ref{logdetb}) and (\ref{logdetf}) leads to an interesting cancellation of the $\sim\zeta(3)$ terms.\footnote{In ABJM theory a similar cancellation \cite{Drukker:2010nc} between the fermionic and bosonic functional determinant of a supersymmetric matter multiplet leads to $|\mathcal{Z}_{\mathrm{matter}}^{\mathrm{1-loop}} (S^3)|^2= 2^{-{{d}_{\mathrm{R}}}}$ where $d_\mathrm{R}>0$ is the dimension of a given representation of the gauge group. Due to the switched statistics of the fields in our $\mathcal{N}=2$ superconformal theory, we find an integer $2^N$ for the square of the $S^3$ partition function.} Implementing this into (\ref{eq:HS_S4_N2}) yields
\begin{equation}\label{SUSYlargeN}
     \log  \mathcal{Z}_{\mathcal{N}=2\,\mathrm{h.s.}}^{(N)}[S^4] \approx  N \left( \log 2-\frac{3\pi}{8\varepsilon_+}-\frac{3\pi}{8\varepsilon_-}\right)~.
\end{equation}
There is still an $\mathcal{O}\left(\tfrac{1}{\varepsilon}\right)$ divergence that can be reabsorbed into a local counterterm, and we distinguish the cutoff from each of the two $\mathcal{Z}_{\mathrm{free}}^{(-N)}[\mathcal{B}_s]$ in (\ref{eq:HS_S4_N2}) by a subscript $\{\pm\}$. Alternatively,  $\varepsilon$ could be taken to be complex valued due to the compexified contour of the bulk conformal factor. It is tempting to take $\varepsilon_+ = - \varepsilon_-$, such that the divergence cancels. Although we do not know the value of the bulk on-shell action on the $S^4$ saddle, expression (\ref{SUSYlargeN}) appears to be compatible with the previous discussion leading to (\ref{SUSYdS}).

\subsubsection*{Curious one-loop cancellations}

At one loop, following the discussion of the previous section, we must compute a Gaussian path integral over the linearised sources, yielding the supersymmetric extension of (\ref{chs2}), namely
\begin{equation}\label{eq:oneloop N=2}
\mathcal{Z}_{\mathcal{N}=2\,\mathrm{h.s.}}^{(N)}[S^4]\approx 2^{N} \times \frac{(-i)^{\mathcal{P}}}{\mathrm{vol}_{{N}}\mathcal{G}_{\mathrm{sHS}}}  \times   \mathcal{Z}_{1-\mathrm{loop}~\text{sHS}}~.
\end{equation}
The partition function $\mathcal{Z}_{\mathcal{N}=2\,\mathrm{h.s.}}^{(N)}$ exhibits a series of curious cancellations/features which we will now list. In fact, we already saw the first of these, namely that the $\sim\zeta(3)$ from the fermionic and bosonic determinants cancel.
\paragraph{Conformal higher spin fields}
We must  compute the supersymmetric conformal higher spin field path integral in (\ref{eq:oneloop N=2}).  At one-loop, we have 
\begin{equation}
\mathcal{Z}_{1-\mathrm{loop}~\text{sHS}}  \equiv \prod_{s\in \mathrm{integer}}{Z}^{(s)}_{1-\mathrm{loop~sHS}}  \prod_{s\in \text{half-integer}}  \mathsf{Z}^{(s)}_{1-\mathrm{loop~sHS}}~.
\end{equation}
The one-loop structure of the conformal higher spin gauge fields, again follows from the general expression (\ref{chschi}). In order to implement (\ref{chschi}), we need to know the bulk and edge characters of totally massless higher spin gauge fields in dS$_4$ and AdS$_4$. For bosonic fields, this was accomplished in \cite{Anninos:2020hfj,Sun:2020ame}. For fermionic fields in dS$_4$, a general expression was argued in equation (6.31) of \cite{Anninos:2025mje}. In appendix \ref{app:fermionic}, we present the results for fermionic higher spin fields in AdS$_4$. 
\newline\newline
The result is quite remarkable. The one-loop characters of the super-conformal higher spin fields for all integer and half integer spins $s\geq \tfrac{1}{2}$ vanish. For $s=\tfrac{1}{2}$ this vanishes because the conformal source has weight $\Delta=\tfrac{3}{2}$ which is equal to its own shadow weight, such that (\ref{chschi}) yields a vanishing result. For higher spins the vanishing is a manifestation of the absence of locally propagating degrees of freedom in three dimensions, though there remains the group volume contribution. Finally, the only non-vanishing characters are at $s=0$. However, the two conformal scalar characters are associated to sources for operators of conformal weights $\Delta=1$ and $\Delta=2$. Implementing (\ref{chschi}) we see that the associated characters differ only by an overall sign and hence their contribution, once again, cancels. Thus,
\begin{equation}
  \log \mathcal{Z}_{1-\mathrm{loop}~\text{sHS}}= 0~.
\end{equation}
We summarize these results in table \ref{tab:HS}. 
\begin{table}[ht]
    \centering
    \small
         \setlength{\arrayrulewidth}{1.2pt}
    \renewcommand{\arraystretch}{1.5}
    \begin{tabular}{|c!{\vrule width .8pt}c!{\vrule width .8pt}c|}
    \hline
       $s$ & $\chi^{(s)}_{\mathrm{sHS~bulk}}$ & $\chi^{(s)}_{\mathrm{sHS~edge}}$ \\
        \hline
          \multirow{2}{*}{$0$}
        & $\Delta =1:\quad \frac{-\e^{-\mathfrak{t}}}{(1-\e^{-\mathfrak{t}})^2}$ &  \multirow{2}{*}{$0$}\\
        \hhline{|~|~|~|} 
        & $\Delta =2:\quad \frac{\e^{-\mathfrak{t}}}{(1-\e^{-\mathfrak{t}})^2}$ & \\
        \hline
        $s\geq 1$ & 0 & 0 \\
        \hline
        $s\geq \frac{1}{2}$ & 0 & 0 \\ \hline
    \end{tabular}
    \caption{Bulk and edge characters of conformal higher spin fields on $S^3$.}
    \label{tab:HS}
\end{table}
\paragraph{Supersymmetric Polchinski phase.}
In (\ref{eq:oneloop N=2}), we have denoted by $\mathcal{P}$ the generalization of Polchinski's phase in the one-loop sphere partition function to the supersymmetric higher spin case. Due to the doubling of each higher spin field in the bulk spectrum, it  can be (reasonably) argued that the Polichinski phase $(-i)^{\mathcal{P}}$ vanishes. The basic reason is that each higher spin with integer $s$ contributes either $\{\pm 1 \}$ to the phase, which always squares to one. Fermionic path integrals don't have the problem of Gaussian unsuppressed path integrals. Nonetheless, massless half-integer gauge fields in de Sitter have imaginary `mass', as recently analysed in \cite{Anninos:2025mje}. For the $\mathcal{N}=2$ theory, however, the spectrum of the Dirac operators comes in complex conjugate pairs despite the imaginary fermionic `mass'. Thus, in (\ref{eq:HS_S4_N2}) we have taken $(-i)^{\mathcal{P}}\equiv1$. 

\paragraph{Dimension of higher spin supergroup.}
The dimension of the three-dimensional conformal higher spin supergroup is given by the weighted sum over the number of conformal Killing tensors and spinors $n_{s,\text{CKT}} = \frac{1}{3}s(4s^2-1)$,\footnote{For instance $2\times n_{1,\mathrm{CKT}}=2$ corresponds to the generators of  two $\mathrm{U}(1)$'s, $2\times n_{2,\mathrm{CKT}}=2\times 10$ corresponds to the number of the (doubled) $\mathrm{SO}(1,4)$ generators, and $2 \times n_{3/2,\mathrm{CKT}}=8$ corresponds to the number of fermonic generators in the free $\mathcal{N}=2$ superconformal theory.} namely
\begin{equation}
 \dim  \mathcal{G}_{\mathrm{sHS}} =\lim_{\delta \rightarrow 0^+} \sum_{s=1,\tfrac{3}{2},\ldots}  2(-1)^{2s}\frac{s(4s^2-1)}{3}\e^{-\delta s}= \frac{1}{8}~.
\end{equation}
We note that in stark contrast to the dimension of the bosonic higher spin supergroup, which is strictly infinite, the above weighted sum is finite upon a simple regularisation prescription. 
This, in particular, implies that
\begin{equation}
  \mathrm{vol}_{{N}}\mathcal{G}_{\mathrm{sHS}}= N^{\frac{\dim  \mathcal{G}_{\mathrm{sHS}}}{2}} \mathrm{vol}\mathcal{G}_{\mathrm{sHS}}  ~,
\end{equation}
where the volume on the right hand side no longer depends on $N$. 

\paragraph{Volume of higher spin supergroup.} We must also confront the volume, ${\mathrm{vol}\,\mathcal{G}_{\mathrm{sHS}}}$, of the superconformal higher spin group, or some real form of its complexification. It is not clear that this problem is well-defined. We will proceed somewhat formally by listing some properties that the desired result might exhibit. We are interested in the real form $\mathcal{G}_{\mathrm{sHS}}$ that has a $\mathrm{Spin}(5)\times \mathrm{U}(1)$ bosonic subalgebra. This belongs, in turn, to a particular real form of the complexified $\mathcal{N}=2$ three-dimensional superconformal group. A natural candidate  is the unitary orthosymplectic supergroup $\text{UOSp}(2|4)$ whose properties we describe in appendix \ref{supergroupapp}. As is generic for supergroups, the super-volume of $\text{UOSp}(2|4)$ vanishes due to unsaturated Grassmann integrals. (The corresponding $\mathcal{N}=1$ counterpart, $\text{UOSp}(1|4)$, on the other hand, has non-vanishing super-volume.) This might suggest that the super-volume of $\mathcal{G}_{\mathrm{sHS}}$ itself might vanish also, as it is a quotient of the universal enveloping algebra \cite{Boulanger:2013zza}. However, establishing this will require a careful treatment of what one means precisely by $\mathcal{G}_{\mathrm{sHS}}$, since the exponentiation of higher spin algebras into higher spin groups is not yet on firm mathematical footing \cite{Monnier:2014tfa}. Given the infinite number of generators in $\mathcal{G}_{\mathrm{sHS}}$, a vanishing seed supergroup volume might be a blessing rather than a curse. In any case, this problem remains open and might require adding a local feature, perhaps akin to the supersymmetric Wilson lines \cite{Mikhaylov:2014aoa} (or the methods in \cite{Rozansky:1992td}) constructed in the Chern-Simons theory with a gauged supergroup.

\subsection{$\mathcal{N}=2$ superglubits}

Putting it all together, after the dust settles, the final result up to one-loop order is given by
\begin{equation}\label{eq:oneloop N=2 final}
    \mathcal{Z}_{\mathcal{N}=2\,\mathrm{h.s.}}^{(N)}[S^4]\approx  2^{N} \times\frac{N^{-\frac{1}{16}}}{\mathrm{vol} \, \mathcal{G}_{\mathrm{sHS}}}  ~,
\end{equation}
up to local divergences of the type $\mathcal{O}( {\varepsilon}^{-1})$ that can be removed with counterterms. The volume of $\mathcal{G}_{\mathrm{sHS}}$ in the above expression is independent of $N$. In particular, it will cancel out from the ratio of two $\mathcal{N}=2$ higher spin $S^4$ partition functions with $N$ and $M$ 
\begin{equation}\label{eq:ratio}
   \mathfrak{r}_{N,M} \equiv \frac{\mathcal{Z}_{\mathcal{N}=2\,\mathrm{h.s.}}^{(N)}[S^4]}{\mathcal{Z}_{\mathcal{N}=2\,\mathrm{h.s.}}^{(M)}[S^4]} \approx 2^{N-M}\left(\frac{M}{N}\right)^{\frac{1}{16}}~.
\end{equation}
We see  that, at least to the order that we are working, when $N=n^{16} k$ and $M=m^{16} k$, with $\{m,n,k\}\in\mathbb{Z}^+$, the ratio  (\ref{eq:ratio}) is rational. For instance, when $(k,m,n) = (1,2,432)$ the ratio is given by $\log \mathfrak{r}_{N,M} \approx {10^{42}}$. 
\newline\newline
Perhaps these results offer some clarity on a counting interpretation of the hypothetical de Sitter horizon entropy \cite{Gibbons:1977mu,Banks:2003cg,fischler2000taking,Parikh:2004wh}. It is tempting to view the correction to $2^N$ in (\ref{eq:oneloop N=2 final}) as a type of topological entanglement entropy \cite{Kitaev:2005dm,Levin:2004mi} associated to an edge-entanglement (or quantum dimension) of the static patch \cite{Anninos:2021ihe}.

\section{Outlook}\label{sec6}

We would like to end  with a few relevant open directions. These complement the results we have developed both conceptually and technically.

\subsection*{Beyond one-loop} 

As has already been noted, the space of supersymmetric sources $\mathcal{B}_s$ transforms as a bilocal $\mathcal{N}=2$ supermultiplet. As such, the effective action stemming from the partition function, $S^{\text{eff}}_{\mathcal{N}=2} = -\log\mathcal{Z}_{\mathrm{free}}^{(-N)}[\mathcal{B}_s]$, can itself be viewed as an $\mathcal{N}=2$ supersymmetric theory invariant under the higher spin supersymmetries. Expanding it out to Gaussian order, yields a real quadratic functional of the $\mathcal{B}_s$. (We are assuming a renormalisation scheme where all one-point functions and contact terms vanish.) As such, and assuming the path integration measure preserves all the symmetries, the Gaussian theory can be viewed as a free $\mathcal{N}=2$ invariant action akin to a gauge invariant $D$-term. 
In turn, $D$-terms are generically $Q$-exact for parity invariant three-dimensional $\mathcal{N}=2$ supersymmetric theories. The essential reason for this is that $D$-terms are given by integrals over all superspace coordinates, which translates to acting  with all available supercharges. This point can bear subtleties: the superconformal higher spin gauge theory has gauge redundancies, and is spatially non-local.\footnote{Moreover, one is gauging the supersymmetry in such a way that the supercharge is a gauge charge on a closed manifold which should thus vanish. Nonetheless, there is an effective supercharge once we expand around the three-sphere background. This is analogous to the fact that when we study perturbative gravity around a spacetime with isometries, we can use the isometries to constrain the structure of correlation functions.} None of these obstructions seem insurmountable, but must be dealt with very delicately. If the Gaussian part is indeed $Q$-exact, it would allow for a localisation type approach to the full path integral (\ref{eq:HS_S4_N2}), as reviewed for example in \cite{Willett:2016adv}, potentially rendering it one-loop exact. Irrespective of the $Q$-exactness, if there exists a renormalisation scheme that preserves the conformal higher spin gauge invariance one can expect the loop corrections to be severely restricted. A complete analysis will require a proper definition of the measure $[\mathcal{D}\mathcal{B}_s]$, perhaps adapting the results of \cite{Anninos:2017eib} to this context.

\subsection*{Observers $\&$ quasi-local features}

There has been a recent emphasis on the potential necessity of quasi-local features in order to properly formulate a theory of cosmology, and more specifically one with cosmological horizons. In their simplest form, such features might look like an infinitesimal worldline decorating the spacetime \cite{Anninos:2017hhn,Chandrasekaran:2022cip,Maldacena:2024spf}, provided it is able to host external operators capable of some primitive measurement. More generally, one could imagine a thickened worldline or finite size boundary harbouring a richer host of information along its surface \cite{Anninos:2024wpy,Silverstein:2024xnr}. From a Euclidean perspective, there appears to be no immediate reason to include such a feature, although it has been suggested that the presence of a phase disrupts the entropic interpretation of the sphere path integral and can be amended by the inclusion of such a feature \cite{Maldacena:2024spf}. Nonetheless, such a phase may \cite{Polchinski:1988ua} or may not \cite{Anninos:2021ene,Giombi:2026sqa} be present, depending on the specific theory under consideration. Moreover, expression (\ref{worldline}) is indicative of a maximally mixed state which has been argued to be the  density matrix describing a Lorentzian de Sitter static patch in the presence of a worldline \cite{Chandrasekaran:2022cip}. 
\newline\newline
In any event, it is of interest to consider whether a gluing formula of the type (\ref{eq:HS_S4}) is a defining property of $\Lambda>0$ quantum gravity. In such a case and irrespective of any additional external data, quasi-local features might be necessary (here a distinguished finite-size $S^3$ embedded inside of $S^4$) rather than optional in defining the sphere path integral itself. As emphasised in \cite{Anninos:2021eit}, one is reminded of early attempts to sum over closed manifolds in two-dimensional quantum gravity --- due to the asymptotic nature of the sum, the inclusion of boundaries were  eventually necessitated \cite{polchinski1994combinatorics}. Further to this, we note that the vanishing supergroup volume of $\text{UOSp}(2|4)$ might require us to further decorate our sphere path integral with additional operators that soak up the zero modes, somewhat analogous to picture changing operators of the superstring. (We note again that for the $\mathcal{N}=1$ case, the supergroup volume is non-vanishing.) Perhaps this is a further indication that a sensible sphere path integral, at least for certain theories, must be accompanied by additional features.

\subsection*{The Cosmological Constant problem}

We end with a speculation about  the cosmological constant \cite{Weinberg:1988cp}. Throughout this work, we have noted that the dimensionless combination $\tfrac{1}{G_N\Lambda}$ is related to a positive integer $N$ in a simple way in the completion of the higher spin theory. This means that $\tfrac{1}{G_N\Lambda}$ cannot take arbitrary values, but rather a discretuum of values fixed by the allowed values of $N$. The $S^4$ path-integrals explored in this work, particularly for the supersymmetric case, have a lot of structure. For a more generic theory of $\Lambda>0$ quantum gravity, however, we can at best expect a perturbative quantum mechanical structure of the type
\begin{equation}
\mathcal{Z}_{\text{grav}} = \sum_{n \in \text{saddles}} { \e^{\alpha_n \mathcal{S}_0+ \gamma_n \log \mathcal{S}_0}} \sum_{l\in \mathbb{N}} {\beta_{nl}}\, \mathcal{S}_0^{-l}~,
\end{equation}
where $\mathcal{S}_0$ is a quantity that goes as $\tfrac{1}{G_N\Lambda}$. The partition function, $\mathcal{Z}_{\text{grav}}$, encodes an infinite collection of loop contributions around each relevant   Euclidean saddle. The coefficients $\{\alpha_n,\beta_{nl}\}$ will generally be transcendental numbers fixed by the properties of the number of matter fields, the structure of the interactions, and so on. The $\gamma_n$ are often found to be rational numbers in saddle point calculations \cite{Anninos:2025ltd}. A simple example of this may be coded in the large $N$ expansion of (\ref{eq:HS_S4}). 
\newline\newline
It is conceivable that in a more generic theory, $G_N\Lambda$ is still a function of some integer or takes a discrete set of allowed values. (For instance, in string theory it may be related to certain fluxes that are subject to charge quantization \cite{Bousso:2000xa,Kachru:2003aw}.) If we further require that a consistent theory sets $\mathcal{Z}_{\text{grav}} $ to be integer valued, or that the ratio of two $\mathcal{Z}_{\text{grav}}$ be rational valued, or something else of the like, those values within the discretuum of  $G_N\Lambda$ that are still allowed are severely restricted --- they may take only a sporadic set of large values that have no simple or uniform distribution. Mathematical equations, like the Diophantine equations, whose variables are restricted to discrete sets can often display such properties \cite{amthor1880problema}.

\section*{Acknowledgments}
It is a great pleasure to acknowledge Andreas Blommaert, Frederik Denef, Dami\'an Galante, Mauricio Romo, Leonardo Santilli, Nati Seiberg, and Zimo Sun for useful discussions.
D.A. is deeply indebted to Frederik Denef, with whom many of these ideas were initiated. B.M. gratefully acknowledges funding provided by the Leinweber foundation at the Institute for Advanced Study and the National Science Foundation with grant number PHY-2207584. D.A.\ is funded by the Royal Society under the grant ``Concrete Calculables in Quantum de Sitter,'' the STFC consolidated grant ST/X000753/1, and the KU Leuven grant C16/25/010. C.B. is funded by STFC under grant reference STFC/2887726. The work of V.A.L. is supported by the ULYSSE Incentive Grant for Mobility in
Scientific Research [MISU] F.6003.24, F.R.S.-FNRS, Belgium.

\appendix
\section{Conformal higher spin fields}\label{app:fermionic}

In this appendix we obtain the characters of conformal higher spin fields on $S^3$. To do so we use the relation \cite{Giombi:2013yva,Anninos:2020hfj}
\begin{equation}\label{eq:HS relation}
    \chi_{\mathrm{HS}_d} = \chi_{\mathrm{dS}_{d+1}} - 2\chi_{\mathrm{AdS}_{d+1}^+}~,
\end{equation}
for $d=3$. Below we review the integer spin $s$ case in AdS$_4$, and derive the characters for half-integer. The integer and half-integer spin characters for dS$_4$ have been obtained in \cite{Anninos:2020hfj, Anninos:2025mje}. We report all the characters in table \ref{tab:PMF}.
\newline\newline
The one-loop heat kernel of a free field in AdS$_4$ carrying a unitary irreducible representation of $\mathrm{SO}(2,d)$ with spin $s$ and conformal dimension $\Delta=\frac{3}{2}+\nu$ has been calculated in \cite{Sun:2020ame} and is given by
\begin{equation}\label{eq:Zsnu_AdS4}
    \log \mathcal{Z}_{s,\nu} = \frac{\mathrm{vol}\,\mathrm{AdS}_4}{\mathrm{vol} \, S^4} \frac{D_s^{(3)}}{4} \int_0^\infty \frac{\d \tau}{2\tau} \e^{-\frac{\varepsilon^2}{4\tau}} \int_0^\infty \d\lambda \mu_s(\lambda)\e^{-\tau(\lambda^2 + \nu^2)}~,
\end{equation}
where we have used the regularised volume of AdS$_4$, along with the standard volume of $S^4$
\begin{equation}
    \mathrm{vol}\,\mathrm{AdS}_4 =\frac{4}{3}\pi^2 ~,\quad \mathrm{vol}\,S^4 = 2\pi^2~.
\end{equation}
The dimension of a spin $s$ representation in $\mathfrak{so}(d)$ is
\begin{equation}\label{eq: dimension Dds}
   D^{(d)}_s = \frac{(d+2s-2)\Gamma(d+s-2)}{\Gamma(d-1)\Gamma(s+1)}~. 
\end{equation}
For example, $D^{(3)}_s\equiv 2s+1$ is the degeneracy of a spin $s$ field in four spacetime dimensions. 
A Hubbard-Stratonovich transformation leads to 
\begin{equation}\label{eq:Z_HS_1}
     \log \mathcal{Z}_{s,\nu} = \frac{D_s^{(3)}}{3!}\int_0^\infty \frac{\d\tau}{2\tau}W_s(\tau)\e^{-\nu\tau}~,\quad W_s(\tau)\equiv \int_{-\infty}^\infty {\d\lambda}\mu_s(\lambda) \e^{i\lambda \tau}~,
\end{equation}
where we formally put $\varepsilon=0$ in (\ref{eq:Zsnu_AdS4}). Finally, $\mu_s(\lambda)$ is the spin $s$ spectral density up to normalization \cite{Camporesi:1994ga, Sun:2020ame}.
\subsection{Bosonic higher spin fields in EAdS$_4$}
 For a \textit{bosonic} spin $s$ field the spectral density $\mu^{(\mathrm{bos})}_s(\lambda)$ is given by\footnote{For $\lambda \rightarrow i\lambda$ this is the Plancherel measure on the principal series of $\mathrm{SO}(1,d+1)$.} 
\begin{equation}
    \mu^{(\mathrm{bos})}_s(\lambda)= \left(\lambda^2 + \left(s+\frac{1}{2}\right)^2\right){\lambda}{\tanh(\pi\lambda)}~.
\end{equation}
To evaluate (\ref{eq:Z_HS_1}) with $\mu_s(\lambda) = \mu^{(\mathrm{bos})}_s(\lambda)$ we close the contour in the upper half-plane and pick up the residues of $ \tanh(\pi \lambda)$ at $\lambda = i(k+\frac{1}{2})$, $k\in \mathbb{N}$. This leads to 
\begin{align}\label{eq:Ws_bos}
    W^{(\mathrm{bos})}_{s} (\tau) &= 2\pi i \times \frac{i}{\pi} \sum_{k\geq 0}\Big(k+\tfrac{1}{2}\Big)\left(\Big(s+\tfrac{1}{2}\Big)^2-\Big(k+\tfrac{1}{2}\Big)^2\right)\e^{-(k+\frac{1}{2})\tau}\cr
    &=-\frac{1+\e^{-\tau}}{1-\e^{-\tau}}\frac{\e^{-\frac{\tau}{2}}}{(1-\e^{-\tau})^3} \left((1-\e^{-\tau})^2s(s+1) -6\e^{-\tau}\right)~.
\end{align}
We can now turn to the fields of interest: Partially massless fields (PMF) in AdS$_4$. PMF in AdS$_{d+1}$ with depth $t$ have conformal dimension $\Delta_{s,t} = d+t-1$. In particular for a PMF of depth $t=s-1$, the one-loop heat kernel partition function (\ref{eq:Z_HS}) needs to account for the gauge redundancy. It is thus given by 
\begin{equation}\label{eq:Zst_bos}
    \log \mathcal{Z}^{\mathrm{bos}}_{s,s-\frac{1}{2}}= \frac{1}{3!} \int_0^\infty \frac{\d \tau}{2\tau} \left(D^{(3)}_s W^{(\mathrm{bos})}_s(\tau)\e^{-(\frac{1}{2}+t)\tau} - D^{(3)}_t W^{(\mathrm{bos})}_t(\tau) \e^{-(\frac{1}{2} +s)\tau} \right)~.
\end{equation}
The first part is the spin $s$ partition function; the second part in the integral subtracts the gauge redundancy. Explicitly, a PMF of depth $t$ has the gauge redundancy 
\begin{align}
    \delta\phi_{\mu_1 \ldots \mu_s}  &= \nabla_{(\mu_{t+1}\ldots\mu_s} \zeta_{\mu_1 \ldots \mu_t)} + \ldots~\cr
    \Delta_- = 1-t &\overset{s-t}{\longleftarrow} \Delta_- = 1-s~,\cr
    \Delta_+ = d+t-1 &\longleftarrow \Delta_+ = d+s-1~.
\end{align}
$\Delta_{\pm}$ are the conformal dimension and the shadow conformal dimension of the boundary fields. Acting with $\nabla_{\mu_{t+1}\ldots\mu_s}$ on $\zeta_{\mu_1 \ldots \mu_t}$ increases the dimension by $(s-t)$ hence we can read off $\Delta_-$ and $\Delta_+$ of $\zeta_{\mu_1 \ldots \mu_t}$. In (\ref{eq:Zst_bos}) we report the partition function with $+$ boundary conditions.
\newline\newline
We note that $D^{(3)}_s- D^{(3)}_{t}$ is exactly the number of degrees of freedom of a PMF at depth $t$.
Combining (\ref{eq:Zst_bos}) and (\ref{eq:Ws_bos}) we find
\begin{equation}
    \log \mathcal{Z}^{\mathrm{bos}}_{s,s-\frac{1}{2}} = \int_0^\infty \frac{\d\tau}{2\tau} \frac{1+\e^{-\tau}}{1-\e^{-\tau}} \left(\chi^{(s)}_{\mathrm{bulk}}(\tau) - \chi^{(s)}_{\mathrm{edge}}(\tau)\right)~,
\end{equation}
where 
\begin{subequations}\label{eq:AdS_boson_characters}
    \begin{align}
       \chi^{(s)}_{\mathrm{bulk}}(\tau) &= \frac{D_s^{(3)}\e^{-(s+1)\tau} -D_{s-1}^{(3)}\e^{-(s+2)\tau}}{(1-\e^{-\tau})^3}~,\\ 
       \chi^{(s)}_{\mathrm{edge}}(\tau)&= \frac{D_{s-1}^{(5)}\e^{-s\tau} -D_{s-2}^{(5)}\e^{-(s+1)\tau}}{(1-\e^{-\tau})}~.
    \end{align}
\end{subequations}
$\chi^{(s)}_{\mathrm{bulk}}(\tau)$ is the character of the $\mathrm{SO}(2,3)$ AdS$_4$ isometry group with $+$ boundary conditions (eq. 9.7 in \cite{Anninos:2020hfj}), whereas $\chi^{(s)}_{\mathrm{edge}}(\tau)$ is the co-dimension two edge contribution.

\subsection{Fermionic higher spin fields in AdS$_4$}
For a \textit{fermionic} spin $s$ field the spectral density in (\ref{eq:Zsnu_AdS4}) is instead given by
\begin{equation}
    \mu^{(\mathrm{fer})}_s(\lambda)= \left(\lambda^2 + \left(s+\frac{1}{2}\right)^2\right)\frac{\lambda}{\tanh(\pi\lambda)}~.
\end{equation}
For the fermionic case, we need to evaluate 
\begin{equation}\label{eq:Z_HS}
     \log \mathcal{Z}_{s,\nu} = -\frac{D_s^{(3)}}{3!}\int_0^\infty \frac{\d\tau}{\tau}W^{(\mathrm{fer})}_s(\tau)\e^{-\nu\tau}~,\quad W^{(\mathrm{fer})}_s(\tau)\equiv \int_{-\infty}^\infty {\d\lambda}\mu^{(\mathrm{fer})}_s(\lambda) \e^{i\lambda \tau}~,
\end{equation}
where now $s$ is a half-integer. We evaluate $W^{(\mathrm{fer})}_s$ by picking up the residues at $\lambda = ik$, $k\in \mathbb{N}$. This leads to 
\begin{align}\label{eq:Wfermion}
    W^{(\mathrm{fer})}_s(\tau)&= 2\pi i\times \frac{i}{\pi} \sum_{k=0}^\infty k\left(\Big(s+\tfrac{1}{2}\Big)^2-k^2\right)\e^{-k\tau}~\cr
    &= \frac{\e^{-\frac{\tau}{2}}}{1-\e^{-\tau}}\frac{\e^{-\frac{3\tau}{2}}}{(1-\e^{-\tau})^3} \left(9+4s(1+s) +\frac{1}{2}\e^{\tau} (1+\e^{-2\tau})(3-4s(s+1))\right)~.
\end{align}
\paragraph{Spin $s=1/2$ and $\Delta = 3/2$.}
We start with the Dirac fermion in AdS$_4$ which has $\nu=0$ in (\ref{eq:Zsnu_AdS4}) and $D_{\frac{1}{2}}^{(3)}=2$:
\begin{eqnarray}
     \log \mathcal{Z}^{\mathrm{fer}}_{\frac{1}{2},0} = -\frac{2}{3!} \int_0^\infty \frac{\d \tau}{\tau}W^{(\mathrm{fer})}_{\frac{1}{2}}(\tau)~,\quad  W^{(\mathrm{fer})}_{\frac{1}{2}} (\tau) = \int_{-\infty}^{\infty} \d\lambda \mu^{(\mathrm{fer})}_{\frac{1}{2}}(\lambda)\e^{i\lambda\tau}~.
\end{eqnarray}
Using (\ref{eq:Wfermion}) we obtain
\begin{equation}
  W^{(\mathrm{fer})}_{\frac{1}{2}} (\tau) = 2 \sum_{k\geq 1} k(k^2-1)\e^{-n \tau} = 3!\frac{2\e^{-2\tau}}{(1-\e^{-\tau})^4}~,
\end{equation}
and hence
\begin{equation}
   \log \mathcal{Z}^{\mathrm{fer}}_{\frac{1}{2},0} =  \int_0^\infty \frac{\d \tau}{\tau}\frac{-\e^{-\frac{\tau}{2}}}{(1-\e^{-\tau})} \chi_{\mathrm{bulk}}^{(\frac{1}{2})}(\tau)~,\quad \chi_{\mathrm{bulk}}^{(\frac{1}{2})}(\tau) = 4\frac{\e^{-\frac{3\tau}{2}}}{(1-\e^{-\tau})^3}~.
\end{equation}
Note that the Dirac fermion has no edge contribution.
\paragraph{Spin $s$ and $\Delta = 1+s$.}
For the general totally massless spin $s$ case, similarly to the bosonic case (\ref{eq:Zst_bos}), we must subtract the spin $t=s-1$ fermion, leading to
\begin{equation}\label{eq:Zfer}
    \log \mathcal{Z}^{\mathrm{fer}}_{s,s-\frac{1}{2}}= -\frac{1}{3!} \int_0^\infty \frac{\d \tau}{\tau} \left(D^{(3)}_s W^{(\mathrm{fer})}_s(\tau)\e^{-(\frac{1}{2}+t)\tau} - D^{(3)}_t W^{(\mathrm{fer})}_t(\tau) \e^{-(\frac{1}{2} +s)\tau} \right)~.
\end{equation}
Combining (\ref{eq:Zfer}) with (\ref{eq:Wfermion}) we obtain 
\begin{equation}
    \log \mathcal{Z}^{\mathrm{fer}}_{s,s-\frac{1}{2}} = \int_0^\infty \frac{\d\tau}{\tau} \frac{-\e^{-\frac{\tau}{2}}}{1-\e^{-\tau}} \left(\chi^{(s)}_{\mathrm{bulk}}(\tau) - \chi^{(s)}_{\mathrm{edge}}(\tau)\right)~,
\end{equation}
where 
\begin{subequations}\label{eq:AdS_fermion_characters}
    \begin{align}
       \chi^{(s)}_{\mathrm{bulk}}(\tau) &= 2 \frac{(2s+1)\e^{-(s+1)\tau} -(2s-1)\e^{-(s+2)\tau}}{(1-\e^{-\tau})^3}~, \\ 
       \chi^{(s)}_{\mathrm{edge}}(\tau)&= \frac{2}{3} \frac{(s-\tfrac{1}{2})(s+\tfrac{1}{2})(s+\tfrac{3}{2}) \e^{-s\tau}- (s-\tfrac{3}{2})(s-\tfrac{1}{2})(s+\tfrac{1}{2})\e^{-(s+1)\tau}}{(1-\e^{-\tau})}~.
    \end{align}
\end{subequations}
In table \ref{tab:PMF} we summarize the characters (\ref{eq:AdS_boson_characters}) and (\ref{eq:AdS_fermion_characters}) together with the dS$_4$ counterparts derived in \cite{Anninos:2020hfj, Anninos:2025mje}.
\begin{table}[ht]
    \centering
    \small
         \setlength{\arrayrulewidth}{1.2pt}
    \renewcommand{\arraystretch}{2.5}
    \begin{tabular}{|c!{\vrule width 1.2pt}c!{\vrule width .8pt}c|}
        \hline
        spin & $\chi^{(s)}_{\mathrm{bulk}}$ & $\chi^{(s)}_{\mathrm{edge}}$ \\
        \hhline{|=|=|=|}
        dS$_4$ integer $s$ 
        & $2\frac{(2s+1)\e^{-(s+1)\mathfrak{t}} -(2s-1)\e^{-(s+2)\mathfrak{t}}}{(1-\e^{-\mathfrak{t}})^3}$
        & $\frac{1}{3}\frac{(2s+1)s(s+1)\e^{-s\mathfrak{t}} -(2s-1)s(s-1)\e^{-(s+1)\mathfrak{t}}}{(1-\e^{-\mathfrak{t}})}$ \\
        \hhline{|=|=|=|}
        dS$_4$ half-integer $s$ 
        & $4 \frac{(2s+1)\e^{-(s+1)\mathfrak{t}} -(2s-1)\e^{-(s+2)\mathfrak{t}}}{(1-\e^{-\mathfrak{t}})^3}$ 
        & $\frac{1}{6} \frac{D_s^{(3)}D_{s-1}^{(3)}D_{s+1}^{(3)} \e^{-s\mathfrak{t}}- D_{s-1}^{(3)}D_{s-2}^{(3)}D_{s}^{(3)}\e^{-(s+1)\mathfrak{t}}}{(1-\e^{-\mathfrak{t}})}$\\
        \hhline{|=|=|=|}
        AdS$_4$ integer $s$ 
        & $\frac{(2s+1)\e^{-(s+1)\tau} -(2s-1)\e^{-(s+2)\tau}}{(1-\e^{-\tau})^3}$
        & $\frac{1}{6}\frac{(2s+1)s(s+1)\e^{-s\tau} -(2s-1)s(s-1)\e^{-(s+1)\tau}}{(1-\e^{-\tau})}$  \\
        \hhline{|=|=|=|}
        AdS$_4$ half-integer $s$ 
        & $2 \frac{(2s+1)\e^{-(s+1)\tau} -(2s-1)\e^{-(s+2)\tau}}{(1-\e^{-\tau})^3}$ 
        & $\frac{1}{12} \frac{D_s^{(3)}D_{s-1}^{(3)}D_{s+1}^{(3)} \e^{-s\tau}- D_{s-1}^{(3)}D_{s-2}^{(3)}D_{s}^{(3)}\e^{-(s+1)\tau}}{(1-\e^{-\tau})}$ \\
        \hline
    \end{tabular}
    \caption{Highest depth partially massless integer and half-integer spin characters. The first two rows correspond to dS$_4$, the last two rows to AdS$_4$. The dimension $D_s^{(d)}$ is given in (\ref{eq: dimension Dds}), and for $d=3$ we have $D_s^{(3)}=2s+1$. }
    \label{tab:PMF}
\end{table}
Combining the entries of table \ref{tab:PMF} and 
\begin{equation}
    \chi_{\mathrm{HS}_d} = \chi_{\mathrm{dS}_{d+1}} - 2\chi_{\mathrm{AdS}_{d+1}^+}~,
\end{equation}
we obtain that the characters of conformal higher spin fields on $S^3$ for integer $s\geq 1$ and half integer spin $s\geq \frac{1}{2}$ vanish as shown in table \ref{tab:HS}. 

\subsection{Spin $s=0$ conformal higher spin fields}
The conformal higher spin characters for spin $s=0$ we obtain from the Harish-Chandra character of the principal series character of dS$_4$ and the character of a massive scalar in AdS$_4$. Explicitly 
\begin{equation}
    \chi_{\mathrm{dS}_4,\nu}(\mathfrak{t}) = \frac{\e^{-(\frac{3}{2}+i\nu)\mathfrak{t}} + \e^{-(\frac{3}{2}-i\nu)\mathfrak{t}}}{(1-\e^{-\mathfrak{t}})^3}~,\quad \chi_{\mathrm{AdS}_4,\Delta_+}(\mathfrak{t}) = \frac{\e^{-\Delta_+ \mathfrak{t}}}{(1-\e^{-\mathfrak{t}})^3}~.
\end{equation}
Using (\ref{eq:HS relation}) we obtain 
\begin{subequations}
\begin{align}
    (\nu,\Delta_+) = \left(\frac{i}{2},1\right): \quad &\chi_{\mathrm{HS}}(\mathfrak{t}) = -\frac{\e^{-\mathfrak{t}}}{(1-\e^{-\mathfrak{t}})^2}~,\\
     (\nu,\Delta_+) = \left(-\frac{i}{2},2\right): \quad &\chi_{\mathrm{HS}}(\mathfrak{t}) = +\frac{\e^{-\mathfrak{t}}}{(1-\e^{-\mathfrak{t}})^2}~.
\end{align}     
\end{subequations}
We can evaluate the HS one-loop partition function explicitly using (\ref{eq:HS relation}) leading to 
\begin{subequations}\label{eq:evalated_HS}
\begin{align}
  \int_\varepsilon \frac{\d \mathfrak{t}}{2\mathfrak{t}}\frac{1+\e^{-\mathfrak{t}}}{1-\e^{-\mathfrak{t}}}\frac{-\e^{-\mathfrak{t}}}{(1-\e^{-\mathfrak{t}})^2} &= +\frac{\zeta(3)}{8\pi^2} +   \mathcal{O}({\varepsilon^{-3},\varepsilon^{-1}})~,\\ 
  \int_\varepsilon \frac{\d \mathfrak{t}}{2\mathfrak{t}}\frac{1+\e^{-\mathfrak{t}}}{1-\e^{-\mathfrak{t}}}\frac{+\e^{-\mathfrak{t}}}{(1-\e^{-\mathfrak{t}})^2} &= -\frac{\zeta(3)}{8\pi^2} +   \mathcal{O}({\varepsilon^{-3},\varepsilon^{-1}})~. 
\end{align}
\end{subequations}
We have indicated the type of ultraviolet divergences that can appear in the above expressions.

\section{Supergroups}\label{supergroupapp}

In this appendix, we mention some properties for the supergroups of interest to the main text. We follow the notation of \cite{Coulembier_2012} (see also \cite{Hasebe:2011uq} for an exposition). The general class of supergroups we are interested in are unitary orthosymplectic supergroups denoted as $\text{UOSp}(n|2m)$, with $n$ and $m$ being positive integers. The bosonic symmetries of $\text{UOSp}(n|2m)$ form the standard group $O(n) \times \text{USp}(2m)$. We are particularly interested in the case $n=2$ and $m=2$, since $O(2)\cong U(1)$ and $\text{USp}(4) \cong \text{Spin}(5)$, which are the natural bosonic symmetries of the $\mathcal{N}=2$ theory on Euclidean dS$_4$  (along with their higher spin extensions). A Euclidean dS$_2$ counterpart, relevant to the $\mathcal{N}=2$ supergravity studied in \cite{Anninos:2023exn,Muhlmann:2025ngz}, is given by $n=2$ and $m=1$ since $\text{USp}(2) \cong \text{Spin}(3)$. The natural $\mathcal{N}=1$ Euclidean dS$_4$ supergroup is $\text{UOSp}(1|4)$, with its two dimensional counterpart being $\text{UOSp}(1|2)$. The relationship to de Sitter is most immediate in two- and four-dimensions, for instance $\text{USp}(6)$ is not isomorphic to $\text{Spin}(7)$. 
\newline\newline
The supergroup $\text{UOSp}(n|2m)$ has bosonic subgroup $O(n)\times \text{USp}(2m)$. The bosonic dimension is $d_B=\tfrac{n(n-1)}{2}+m(2m+1)$, and the fermionic real dimension is $d_F=2mn$. It admits a definition in terms of $(n|2m)\times (n|2m)$ super-matrices as
\begin{equation}
    \text{UOSp}(n|2m) = \left\{ M = \begin{pmatrix}
        E & F \\ G & H
    \end{pmatrix} \in GL(n|2m) ~~\text{s.t. $M$ satisfies ($u$) and ($sp$)} \right\} ~,
\end{equation}
where
\begin{equation}
    (u): \begin{pmatrix}
        E^\dag & G^\dag \\ F^\dag & H^\dag
    \end{pmatrix} \begin{pmatrix}
        E & -G \\ F & H
    \end{pmatrix} = \begin{pmatrix}
        \mathbf{1}_n & 0 \\
        0 & \mathbf{1}_{2m}
    \end{pmatrix} ~,
\end{equation}
and
\begin{equation}
    (sp): \begin{pmatrix}
        E^T & G^T \\ F^T & H^T
    \end{pmatrix}  g \begin{pmatrix}
        E & -G \\ F & H
    \end{pmatrix} = g ~, ~~~ \text{with} ~~~ g = \begin{pmatrix}
        \mathbf{1}_n & 0 \\ 0 & J
    \end{pmatrix} ~~~ \text{and} ~~~ J = \begin{pmatrix}
        0 & -\mathbf{1}_m \\ \mathbf{1}_m & 0
    \end{pmatrix} ~.
\end{equation}
From the above, we deduce that the bosonic supergroup of $\text{UOSp}(n|2m)$ is given by $O(n)\times \text{USp}(2m)$. The dagger operation is given by Hermitian conjugation, whereby complex conjugation is of the second kind and will be later denoted by a bar, and $T$ represents ordinary transposition. An explicit expression for the invariant measure on $\text{UOSp}(n|2m)$ is given in section 7 of \cite{Coulembier_2012}, and reads
\begin{equation} \label{uosp invariant measure}
    \d\mu = \text{det}(\mathbf{1}_n - \theta^T J \theta)^{-1/2} [\d x,\d z|\d\theta,\d\theta^\dag]~.
\end{equation}
In writing the above expression, we have parameterised a generic supermatrix  as 
\begin{equation}
    M = \begin{pmatrix}
        xA & x\theta^T Jz \\ \theta & Bz
    \end{pmatrix} ~,
\end{equation}
where $x\in O(n)$, $z\in \text{USp}(2m)$. Also, $\theta$ is a $2m\times n$ complex Grassmann matrix satisfying $\theta^\dagger=\theta^T J$, while $A \equiv \sqrt{\mathbf{1}_n-\theta^T J\theta}$, and $B\equiv \sqrt{\mathbf{1}_{2m}-\theta \theta^T J}$. 
\newline\newline
As a simple example, we can consider $n=2$ and $m=1$. Expanding (\ref{uosp invariant measure}) one finds
\begin{equation} \label{uosp2 top form}
    \text{det}(\mathbf{1}_n - \theta^T J \theta)^{-1/2} \big|_{\mathrm{UOSp}(2|2)} = 1+ \bar{\theta}_{11}\theta_{11} + \bar{\theta}_{12}\theta_{12} ~,
\end{equation}
where $\bar{\theta} \equiv -J \theta$. We note that the expression does not contain the $\theta$ top form and consequently the integral over the supergroup $\text{UOSp}(2|2)$ super-volume form vanishes. The vanishing is in line with general arguments for the super-volume of supergroups with compact bosonic subgroups (see footnote 9 of \cite{Mikhaylov:2014aoa} for a related discussion). To obtain a non-vanishing result, we need to soak up the remaining Grassmann zero modes. This can be achieved by inserting an appropriate function $f(x,z,\theta)$ in the integral that produces the top form $\bar{\theta}_{11}\theta_{11}\bar{\theta}_{12}\theta_{12}$. A slightly more cumbersome calculation reveals that the volume of $\text{UOSp}(2|4)$ also vanishes:
\begin{align}
\begin{split}
    \det(1 - \theta^T J\theta)^{-1/2}\big|_{\mathrm{UOSp}(2|4)} & = 1 + \bar{\theta}_{11}\theta_{11} + \bar{\theta}_{12}\theta_{12} + \bar{\theta}_{21}\theta_{21} + \bar{\theta}_{22}\theta_{22} \\
    & + 3\bar{\theta}_{11}\theta_{11}\bar{\theta}_{21}\theta_{21} + \bar{\theta}_{12}\theta_{12}\bar{\theta}_{21}\theta_{21} +\bar{\theta}_{11}\theta_{12}\bar{\theta}_{21}\theta_{22} + \bar{\theta}_{12}\theta_{11}\bar{\theta}_{21}\theta_{22} \\
    & + \bar{\theta}_{11}\theta_{12}\bar{\theta}_{22}\theta_{21} + \bar{\theta}_{12}\theta_{11}\bar{\theta}_{22}\theta_{21} + \bar{\theta}_{11}\theta_{11}\bar{\theta}_{22}\theta_{22} + 3\bar{\theta}_{12}\theta_{12}\bar{\theta}_{22}\theta_{22} ~.
\end{split}
\end{align}
It is straightforward to check that the top form $\bar{\theta}_{11}\theta_{11}\bar{\theta}_{12}\theta_{12}\bar{\theta}_{21}\theta_{21}\bar{\theta}_{22}\theta_{22}$ does not appear in the previous expression, and consequently the volume of the supergroup vanishes. Interestingly, the super-volume of the $\mathcal{N}=1$ counterparts, $\text{UOSp}(1|2)$ and $\text{UOSp}(1|4)$, is non-vanishing and we obtain
\begin{subequations}
\begin{align}
    \det(1 - \theta^T J\theta)^{-1/2}\big|_{\mathrm{UOSp}(1|2)} &=1+\bar{\theta}_{1}\theta_{1}~,\\
    \det(1 - \theta^T J\theta)^{-1/2}\big|_{\mathrm{UOSp}(1|4)} &=1+\bar{\theta}_1\theta_1 +\bar{\theta}_2\theta_2 +3\bar{\theta}_1\theta_1\bar{\theta}_2\theta_2 ~.
\end{align}    
\end{subequations}
Hence we find 
\begin{subequations}
\begin{align}
    \mathrm{vol}(\mathrm{UOSp}(1|2)) &= \mathrm{vol}(\mathrm{O}(1)) \, \mathrm{vol}({\mathrm{SU}}(2))~,\\
    \mathrm{vol}(\mathrm{UOSp}(1|4)) &= 3\,\mathrm{vol}(\mathrm{O}(1)) \,\mathrm{vol}(\mathrm{Spin}(5)) ~.
\end{align}    
\end{subequations}

\bibliographystyle{JHEP}
\bibliography{bib}
\end{document}